\documentclass[12pt,preprint]{aastex}

\slugcomment{To be submitted to ApJ}

\shorttitle{YSOVAR:  ONC PMS Eclipsing Binaries}
\shortauthors{Morales-Calder\'on et al.}

\begin{document}


\title{YSOVAR: Six Pre-Main-Sequence Eclipsing Binaries in the Orion Nebula Cluster}


\author{M. Morales-Calder\'on\altaffilmark{1,20}, J. R. Stauffer\altaffilmark{1}, K. G. Stassun\altaffilmark{2,3,4}, F.~J.~Vrba\altaffilmark{5}, L.~Prato\altaffilmark{6}, L.~A.~Hillenbrand\altaffilmark{7}, S.~Terebey\altaffilmark{8}, K. R. Covey\altaffilmark{9},  L.~M.~Rebull\altaffilmark{1}, D. M. Terndrup\altaffilmark{10,11}, R. Gutermuth\altaffilmark{12}, I.~Song\altaffilmark{13}, P.~Plavchan\altaffilmark{14}, J.~M.~Carpenter\altaffilmark{7},  
F. Marchis\altaffilmark{15}, E. V. Garc\'ia\altaffilmark{3}, S. Margheim\altaffilmark{16},  K. L. Luhman\altaffilmark{17,18}, J. Angione\altaffilmark{8}, J. M. Irwin\altaffilmark{19}}



\email{mariamc@cab.inta-csic.es}


\altaffiltext{1}{Spitzer Science Center, California Institute of Technology, 1200 E California Blvd., Pasadena, CA 91125, USA.}
\altaffiltext{2}{Physics and Astronomy Department, Vanderbilt University, 1807 Station B, Nashville, TN 37235, USA.}
\altaffiltext{3}{Fisk University, Department of Physics, Fisk University, 1000 17th Ave. N, Nashville, TN 37208, USA}
\altaffiltext{4}{Department of Physics, Massachusetts Institute of Technology, Cambridge, Massachusetts, 02139, USA}
\altaffiltext{5}{U.S. Naval Observatory, Flagstaff Station, 10391 W. Naval Observatory Road, Flagstaff, AZ 86001-8521, USA.}
\altaffiltext{6}{Lowell Observatory, 1400 West Mars Hill Road, Flagstaff, AZ 86001, USA}
\altaffiltext{7}{Astronomy Department, California Institute of Technology, 1200 E California Blvd., Pasadena, CA 91125, USA.}
\altaffiltext{8}{Dept. of Physics and Astronomy, California State University at Los Angeles, Los Angeles, CA 90032, USA.}
\altaffiltext{9}{Hubble Fellow; Department of Astronomy, Cornell University, 226 Space Sciences Building, Ithaca, NY 14853, USA.}
\altaffiltext{10}{Department of Astronomy, The Ohio State University, 140 West 18th Avenue, Columbus, OH 43210, USA.}
\altaffiltext{11}{National Science Foundation, 4201 Wilson Boulevard, Arlington, VA 22230, USA}
\altaffiltext{12}{Dept. of Astronomy, University of Massachusetts, Amherst, MA  01003, USA}
\altaffiltext{13}{Physics and Astronomy Dept., University of Georgia,Athens, GA 30602-2451, USA}
\altaffiltext{14}{NASA Exoplanet Science Institute, California Institute of Technology, Pasadena, CA 91125, USA}
\altaffiltext{15}{SETI Institute, Carl Sagan Center, 189 N San Bernado Av, Mountain View CA 94043 USA}
\altaffiltext{16}{Gemini Observatory, Southern Operations Center, Casilla 603, La Serena, Chile}
\altaffiltext{17}{Department of Astronomy and Astrophysics, The Pennsylvania
State University, University Park, PA 16802, USA }
\altaffiltext{18}{Center for Exoplanets and Habitable Worlds, The Pennsylvania
State University, University Park, PA 16802, USA}
\altaffiltext{19}{Harvard-Smithsonian Center for Astrophysics, 60 Garden St., Cambridge, MA 02138, USA}
\altaffiltext{20}{Centro de Astrobiolog\'ia (INTA-CSIC); ESAC Campus, P.O. Box 78, E-28691 Villanueva de la Canada, Spain}


\begin{abstract}
Eclipsing binaries (EBs) provide critical laboratories for empirically
testing predictions of theoretical models of stellar structure and
evolution. Pre-main sequence (PMS) EBs are particularly valuable, both
due to their rarity and the highly dynamic nature of PMS evolution, such that a
dense grid of PMS EBs is required to properly calibrate theoretical
PMS models.  Analyzing multi-epoch, multi-color light curves for
$\sim$2400 candidate Orion Nebula Cluster (ONC) members from our Warm
Spitzer Exploration Science Program YSOVAR, we have identified twelve
stars whose light curves show eclipse features.  Four of these 12 EBs
are previously known. Supplementing our light curves with follow-up
optical and near-infrared spectroscopy, we establish two of the
candidates as likely field EBs lying behind the ONC. We confirm the remaining six candidate systems,
however, as newly identified ONC PMS EBs.  These systems
increase the number of known PMS EBs by over 50\%, and include the
highest mass ($\theta^1$ Ori E, for which we provide a complete set of well determined parameters including component masses of 2.807 and 2.797 M$_{\odot}$) and longest period (ISOY J053505.71-052354.1, P$\sim$20 days) PMS
EBs currently known.  In two cases ($\theta^1$ Ori E and ISOY J053526.88-044730.7),
enough photometric and spectroscopic data exists to attempt an orbit
solution and derive the system parameters. For the remaining systems,
we combine our data with literature information to provide a
preliminary characterization sufficient to guide follow-up
investigations of these rare, benchmark systems.
\end{abstract}


\keywords{open clusters and associations: individual (Orion)---stars: pre-main sequence---stars: binaries: eclipsing---stars: variables: general}



\section{Introduction}

The Orion Nebula Cluster (ONC) contains several thousand members, and since it is nearby, it provides an excellent empirical laboratory to study the physical properties of pre-main sequence (PMS) stars and brown dwarfs.
The ONC is particularly useful for comparison of the observed luminosities and effective temperatures of PMS stars to theoretical model predictions \citep{H97,DaRio10}.  Such comparisons can, in theory, allow an estimate of the masses of individual stars as well as both the mean age and the age spread for the stars in a cluster.  For such estimates to be meaningful, however, the theoretical tracks and isochrones must be vetted against observations to insure that they are well calibrated.  Empirical measurements of the masses, radii, and temperatures of stars, over a range of masses, are necessary for understanding stellar evolution and for deriving well-calibrated theoretical models. 

The most rigorous means to measure precise stellar properties is via
identification and characterization of PMS eclipsing binary (EB) stars because, through a complete analysis of spectroscopy and photometry of these systems, the individual masses, radii, temperatures, and absolute luminosities of the two stars
can be accurately derived. However,  the identification of such systems is difficult due to the need for monitoring observations and the fact that the system must have an inclination close to 90$^\circ$ to be detected. The paucity of known PMS EBs has meant that the theoretical models lack rigorous empirical confirmation, and thus that masses derived from those tracks
have significant systematic uncertainties associated with them \citep{H04}.  The situation
has begun to change recently with the advent of sensitive wide-field cameras,
robotic telescopes and automated photometry pipelines, allowing deep, wide, long
duration time series monitoring programs to be conducted.  These programs have
now led to the identification of seven low-mass PMS EBs with individual masses lower than 1.5 M$_\sun$: RXJ 0529.4+0041A \citep{Covino00}, V1174 Ori \citep{Stassun04}, 2MJ0535-05 \citep{Stassun06, Stassun07}, JW 380 \citep{Irwin07}, Par 1802 \citep{Cargile08, Stassun08}, ASAS J0528+03 \citep{Stempels08}, and MML 53 \citep{Hebb10, Hebb11}. These systems have components ranging in mass from 0.036 M$_\sun$ (2MJ0535-05B) to 1.38 M$_\sun$ (ASAS J0528+03A).  All but ASAS J0528+03,  MML 53, and RXJ 0529.4+0041 are associated with the ONC. There are just another  four EB systems where either only the secondary is on the PMS or both components are more massive PMS stars: EK Cep \citep{Popper87}, TY CrA \citep{Casey98}, RS Cha \citep{Alecian05,Alecian07} and perhaps also V578 Mon \citep{Garcia11} although its components are B type stars that may already be on the main sequence. The masses of a handful PMS low-mass stars have been measured using other methods (see \citet{H04, Boden05,Simon00,Tognelli11} for a summary on dynamical mass determination and calibration of PMS tracks).

Despite the recent discoveries, it is still important to search
for additional PMS EBs.   Main sequence solar-type stars are well described by state-of-the-art stellar evolution models (i.e., observations agree well with theoretical isochrones); however, recent measurements of the stellar properties of low-mass dwarfs and young PMS stars remain problematic for the existing models. Recent work has suggested that
in addition to mass and age, other parameters, such as magnetic field strength or
rotation, may be necessary to fully characterize young, low mass stars. Magnetic fields in young, rapidly rotating low mass stars are thought to inhibit convection and thereby cause those stars to
have larger radii and cooler temperatures than would otherwise be the case \citep{MoralesJC10, Macdonald10}.  This effect has been invoked to explain the properties of the 1 Myr old brown dwarf EB in Orion (2M0535-05) where the more 
massive component is unexpectedly cooler than its companion \citep{Reiners07,Chabrier07,Stassun07,Mohanty10}. 
The effects of magnetic fields on stellar structure are not included in the models 
and are not completely understood \citep{Chabrier07}. Most PMS eclipsing
binaries discovered to date have short enough orbital periods that their
components are expected to have their rotation periods tidally locked to
the orbital period.  Therefore, they are likely to be rapidly rotating and
have strong magnetic fields, and hence 
have inflated radii (see \citet{Kraus11}).  Identification of PMS EBs with longer
periods, where tidal locking is not expected, could offer a direct test of
the proposed link between magnetic fields, rotation and radii. Moreover, 2M0535-05 remains the only known brown dwarf EB.

In this paper, we report the identification and initial characterization of six new PMS EB candidates in the ONC, discovered as part of the YSOVAR (Young Stellar Object VARiability) Spitzer
Exploration Science program \citep{Morales11}.  These systems have been overlooked in the past probably because the area surrounding the Trapezium stars is filled with bright nebulosity, making the optical photometry very unreliable, but also because cadence and duration of observations in some previous studies were not ideal. Details of the discovery observations 
are reported in Section 2.   Follow-up observations 
are presented in Section 3 and 4.   In section 5 we provide a description of the available data and
preliminary analysis for the 6 PMS EBs, based in most cases upon the light curves, in order to provide a basic initial characterization. Two additional systems are likely field star EBs  lying behind the ONC and are described in the Appendix. 

\section{Discovery Light Curves: The 2009 Campaign}

\subsection{Spitzer Observations in 2009}

As part of the Spitzer Exploration Science program YSOVAR, we used about 250 hours of warm Spitzer Space Telescope \citep{Werner04} observing time to monitor $\sim$0.9 deg$^2$ of the ONC with the Infrared Array Camera \citep[IRAC;][] {Fazio04} at 3.6 and 4.5 $\mu$m in Fall 2009. The observed area was broken into five segments (five astronomical observation requests or AORs) with a central region of 
$\sim$20$\arcmin\times$25$\arcmin$ and four flanking fields. The central 
part was observed in full array mode with 1.2 seconds exposure time at 
20 dithering positions to avoid saturation by the bright nebulosity 
around the Trapezium stars. The remaining four segments of the map were 
observed in High Dynamic Range mode with exposure times of 10.4 and 0.4 
seconds, at 4 dithering positions. 
These observations were taken for 40 days in Fall 2009, with $\sim$2 epochs 
each day. We used the IDL package Cluster Grinder \citep{Gutermuth09} which, 
starting from the basic calibrated data (BCD) released by the Spitzer 
Science Center, builds the combined mosaic for each AOR at each epoch and performs aperture photometry on the mosaics. 
Because we dithered while mapping and then extracted the photometry from the
mosaiced image, each star at each epoch will have contributions to its image from four to as many
as 80 BCDs; the time and magnitudes we report for each epoch correspond to the average for all of those frames. 

From these data, we constructed light curves for $\sim$10000 point
sources in our FOV.  About 2400 of these stars were identified as probable
ONC members, including $\sim$1400 stars with probable mid-IR excesses (Megeath
et al. 2012) and $\sim$1000 additional stars identified in the literature as
probable ONC members from X-ray, proper motion, radial velocity and variability studies
\citep{Parenago54,Jones88,McNamara76,Tian96,Tobin09,Carpenter01,Getman05}, but 
lacking IR excess significant enough to be included in the Megeath et 
al. catalog of YSO candidates.   
Most of these latter stars should be weak-lined T~Tauris (WTTs), though some 
may be sources with excesses that escaped previous detection. While our primary goal was to use these data to investigate the structure of the inner disk and time-variable accretion in YSOs with circumstellar disks, these observations also provided a treasure of data on all types of PMS variability. Further discussion of this dataset and initial results can be found in \citet{Morales11}.  

In order to search for eclipsing systems, we first ran a Box-Least-Squares \citep[BLS;][]{Kovacs02} periodogram on all 2400 of these sources and then automatically scanned the folded light curves looking for the signature of detached EBs - short duration flux dips that are present and similar in both IRAC channels and light curves that are otherwise approximately constant. In addition, we also visually inspected the light curves searching for 
signatures that our code could have missed. We also examined in a similar 
fashion the light curves of the $\sim$1000 anonymous stars in our catalog brighter than [4.5] = 13.5 mag (which corresponds to the quiescent magnitude of our faintest EB candidate).

From this process, we identified 12 EB candidates. Four of them, 2MASS 05352184-0546085 \citep{Stassun06}, V1174 Ori \citep{Stassun04}, Par1802 \citep{Cargile08,Stassun08}, and  JW 380 \citep{Irwin07} are already known PMS EBs. 
Another six systems are newly identified ONC PMS EBs --including $\theta^1$ Ori E, a known PMS spectroscopic binary \citep{Costero06,Herbig06} which was flagged as potentially eclipsing, but no confirmatory photometry had been obtained until now-- and are shown in detail in the following sections. The remaining two sources are not known Orion members from the literature and, based on our analysis of the available data, we believe are likely to be background stars, though still eclipsing binaries (see Appendix A1 and A2). All but one (ISOY J053454.31-045413.0, hereafter: ISOY J0534-0454) of the new PMS EB seem to have no detectable IR excess and they are likely to be WTTs.

\begin{figure}
\epsscale{.6}
\plotone{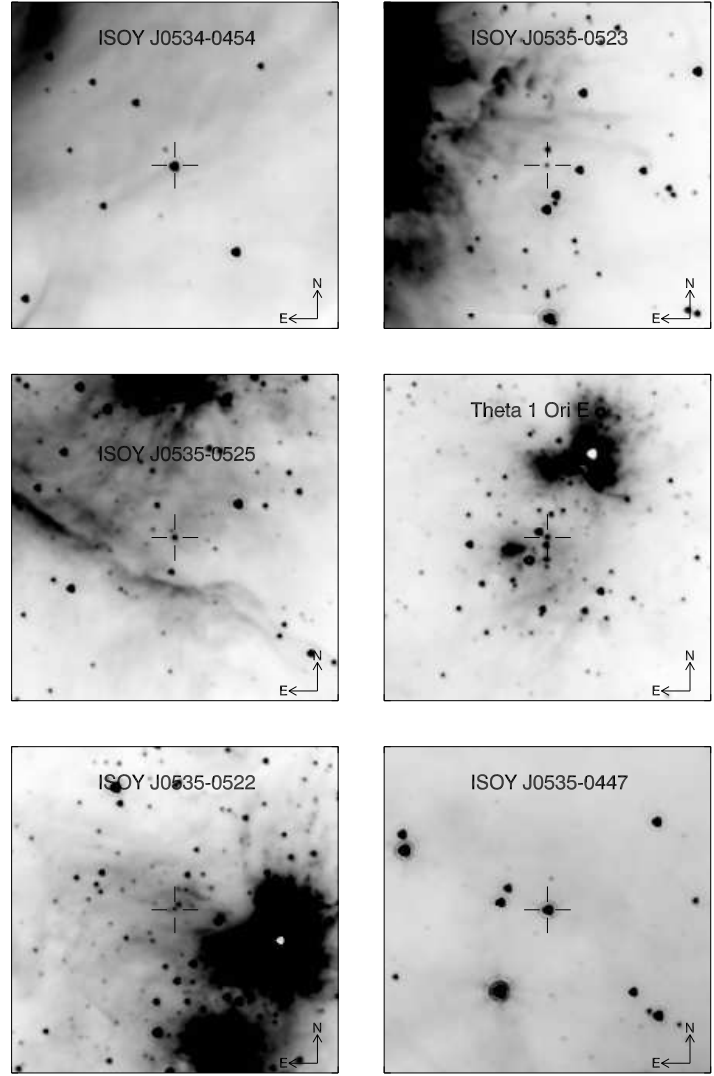}
\caption{Finding charts from the Spitzer 4.5$\mu$m\ images for
each of our 6 PMS EBs ordered by right ascension. The size of each image stamp is 3$\arcmin\times$3$\arcmin$. Note that the intensity scale varies from one image to the next. \label{EB_FC}}
\end{figure}

We used implementation of the BLS period-finding algorithm available at the NExScI site
(http://exoplanetarchive.ipac.caltech.edu) to determine if the flux dips in each light curve are indeed periodic and to identify the best period. In some cases, when we had several bandpasses or epochs well-separated in time we tweaked the obtained period by hand to obtain a more accurate one consistent with all epochs and bandpasses. For the four previously known systems, our derived periods agree, within errors, with the published period. Our new PMS EB candidates are listed
in Table \ref{binaries}. The first column of Table \ref{binaries} displays the name of the source which is formed by an acronym (ISOY: Initial Spitzer Orion YSO) followed by the coordinates of the source. For the sake of simplicity we will use reduced names (i.e.: ISOY JHHMM-DDMM) throughout the text and figures except for ISOY J053515.76-052309.9, for which we will use its most common name in the literature: $\theta^1$ Ori E. The remaining columns in Table \ref{binaries}  show  available broad band photometry, spectral types, and periods.  Finding charts from the Spitzer 4.5~$\mu$m\ images for our 6 PMS EBs are provided in Figure \ref{EB_FC}.

\subsection{Ground-based Observations in 2009}

To complement our Spitzer data, we obtained contemporaneous $I_c$ and $J$ photometry, usually for smaller areas within the Spitzer mosaic. The main source of $J$-band monitoring data was the United Kingdom Infrared Telescope Wide Field Camera (UKIRT/WFCAM). For the $I_c$ band, the New Mexico State University/Apache Point Observatory (NMSU/APO) 1m telescope and NasaCam at the 31$\arcsec$ telescope at Lowell Observatory contributed the bulk of the monitoring data.  We performed differential aperture photometry on the ground-based data. In each dataset, for each object, an artificial reference level was created by selecting 30 nearby isolated stars; we iteratively eliminated those with larger photometric scatter  or evidence of variability. Zero-points for the photometry were established by reference to data
for non-variable stars in our FOV from the Two Micron All-Sky Survey (2MASS) Point Source Catalog \citep{Cutri03} for $J$, and by reference to data in \citet{H97} for $I_c$.

All of the 2009 time series data are publicly available through the YSOVAR database (http://ysovar.ipac.caltech.edu/). The 2009 light curves for the six new PMS EB are shown in Figure~\ref{PMSEBs_allLC2009}.  Where we have photometry at the right epoch, the ground-based data corroborate the Spitzer eclipses.

\section{Follow-up Light Curves: The 2010--2011 Campaign}

To confirm the identification of the new EBs and to refine measurements of their orbital periods and eclipse shapes, we obtained additional photometry in 2010 and 2011. 
These observations are described in the next sections and summarized in Table~\ref{ObsSummary}.

\subsection{Spitzer Observations in 2010 and 2011}

Four of our new PMS EB candidates: $\theta^1$ Ori E, ISOY J0535-0525, ISOY J0535-0523, and ISOY J0535-0522, are located within 3$\arcmin$ of $\theta^1$ Ori C, the brightest star in the Trapezium cluster.  These stars were observed in IRAC full-frame mapping mode with a frametime of 2 seconds and 20 small dithers, at a cadence of 7 to 8 times per day, for the period Nov. 19-29, 2010. Another two systems,  ISOY J0535-0447 and ISOY J0536-0500, were observed for just two days in November (Nov. 20-22) 2010 at the same cadence, but with a 12 second frametime 
 and four small dithers.  The observations were designed to detect at least one eclipse (either primary or secondary) in each system. In addition, sparser data ($\sim$8 epochs in 30 days) was also obtained for all of our EB candidates in Fall 2010 with the same setup as in the 2009 mapping observations (same spatial coverage and integration times). Finally, in November 2011 we obtained three observations per day for ISOY J0536-0500 and ISOY J0534-0454 for about 20 days with a 12 second frametime and four dithers at each epoch. The IRAC light curves for 2010 and 2011 were extracted in the same way as for the 2009 data.  Figure~\ref{PMSEBs_allLC2010} shows these light curves for the 6 PMS EBs. 

For ISOY J0535-0447,  we also obtained $\sim$10 hrs of IRAC staring mode observations at [4.5] on 2011 April 27, scheduled to cover the secondary eclipse of ISOY J0535-0447. In this case
we simply stared at the target object, with no dithering, in order to provide more accurate relative photometry. A total of 2713 consecutive frames of data with 10.4 sec integration time were taken. The integration time was selected so that the number of electrons would be high; however, to make sure that the target would not saturate, we placed it in the center of four pixels. For the light curve extraction of these data, we performed aperture photometry on the corrected BCDs (cBCDs) supplied by the Spitzer Science Center. The cBCDs are calibrated frames with empirical corrections for artifacts due to bright sources. However, instrumental effects such as the pixel phase effect can still remain \citep{Morales06,Cody11}. We inspected our light curve searching for signs of this instrumental signature (flux strongly correlated with sub-pixel position) but, if present, its effect on the light curve is negligible compared to the amplitude of the observed variations. 

\subsection{Ground-Based Observations in 2010 and 2011}
As in 2009, we also obtained contemporaneous $I_c$  and $J$ photometry in 2010 and 2011. The data were obtained from the UKIRT/WFCAM, the NMSU/APO 1m telescope, the Wide Field Imager on the 2.2 m telescope at La Silla Observatory, and the 40$\arcsec$ telescope at the United States Naval Observatory (USNO) Flagstaff Station. The $J$ band UKIRT mosaic covers the whole area observed by Spitzer, and we obtained one observation per night for about 30 days in Fall 2010 (between Oct. 29 and Dec. 15). For 10 of those nights  two observations per night were obtained. APO, WFI and USNO data were obtained for the Trapezium area with a cadence of once a night between October 7 and December 14 for APO, between November 21 and November 28 for WFI, and for several hours a night for 34 nights between October 30 and December 28 for USNO. Finally, the largest effort was performed on the Northern area where ISOY J0535-0447 and ISOY J0534-0449 are located. We used the USNO 40$\arcsec$ telescope to monitor a $\sim$23$\arcmin\times$23$\arcmin$ region for several hours per night for 38 nights in 2010 and 35 nights in 2011 (a total of 15137 frames were obtained). The light curves were extracted in a similar fashion as for the 2009 data. 

\begin{figure}
\epsscale{.7}
\plotone{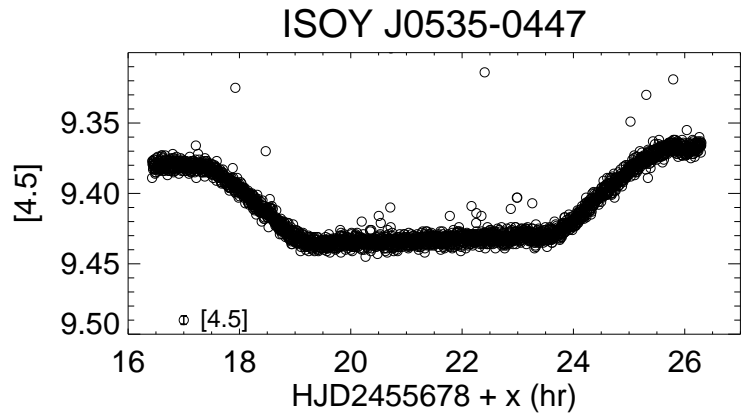}
\caption{Follow-up IRAC [4.5] staring mode light curve for ISOY J0535-0447 obtained on April 27, 2011 showing the secondary eclipse. \label{EB3227_allLC2010}}
\end{figure}

The light curves for our 6 PMS EBs with the follow-up photometry from 2010 and 2011 can be seen in Figures~\ref{PMSEBs_allLC2010} and \ref{EB3227_allLC2010}. All the YSOVAR time series photometry at all available bandpasses for the 6 PMS EBs can be found in Tables~\ref{EB45099_timeseries} to~\ref{EB40003_timeseriesI1}.

\section{Spectroscopic Follow-up}

\subsection{Low-Resolution Spectroscopy}

We obtained moderate-resolution near-infrared spectra of three of our
PMS EB candidates using Triplespec at the Palomar 5m telescope \citep{Wilson04,Herter08}. TripleSpec
covers the wavelength range from 1 to 2.45 $\mu$m  simultaneously at a spectral resolution of $\sim$2700. The entrance slit is 1$\times$30 arcsecond and the spectrum is spread over five orders. Observations were obtained at two slit positions, using a standard ABBA nod sequence.  Spectra of ISOY J0535-0447, ISOY J0535-0522, and ISOY J0535-0523 with total exposure times of 720, 720 and 1200 seconds, respectively, were obtained on September 20 and 21, 2011; another 720 second spectrum of ISOY J0535-0447 was obtained on November 27, 2010 with the same setup.  

We obtained similar spectra for ISOY J0535-0454 and ISOY J0536-0500 using the TripleSpec spectrograph mounted on the APO ARC3.5 telescope \citep{Wilson04}.  APO TripleSpec instrument is a twin of Palomar TripleSpec and thus covers the same wavelength  range (1 to 2.45 $\mu$m) at similar resolution. Both sources were observed on October 30, 2011 at two slit positions and with a total exposure time of 5280 seconds. The 1$\times$45 arcsecond slit was used  yielding a resolution of $\sim$2300.  

The spectra were reduced using a custom version of the IDL-based
SpeXTool reduction package \citep{Cushing04}, as modified to
process data obtained by the Palomar  and APO TripleSpec instruments.  Sky subtraction was performed by differencing
spectral frames obtained with the target at the A and B slit positions;
visual inspection of the sky subtracted frames identified apparent sky
line artifacts in the sky subtracted 2-dimensional spectra of ISOY
J0535-0523 (primarily the He I line at 1.083 microns) and
numerous lines throughout the $JHK$ bands for ISOY J0535-0522.  These
residual `sky lines' are due to line emission from the
substantial nebulous regions near J0535-0523 and J0535-0522, which can
be easily seen in the 4.5 $\micron$ finding charts presented in Figure~\ref{EB_FC}.

Each target's flattened, sky-subtracted, extracted, and wavelength
calibrated spectrum was corrected for telluric absorption and flux
calibrated using the Interactive Data Language (IDL) XTELLCOR package \citep{Vacca04} and
observations of A0V stars obtained nearby in time ($\delta t <$ 30
min) and airmass ($\delta$ sec z $<$ 0.2) to each target observation.

\begin{figure}
\epsscale{1.}
\plotone{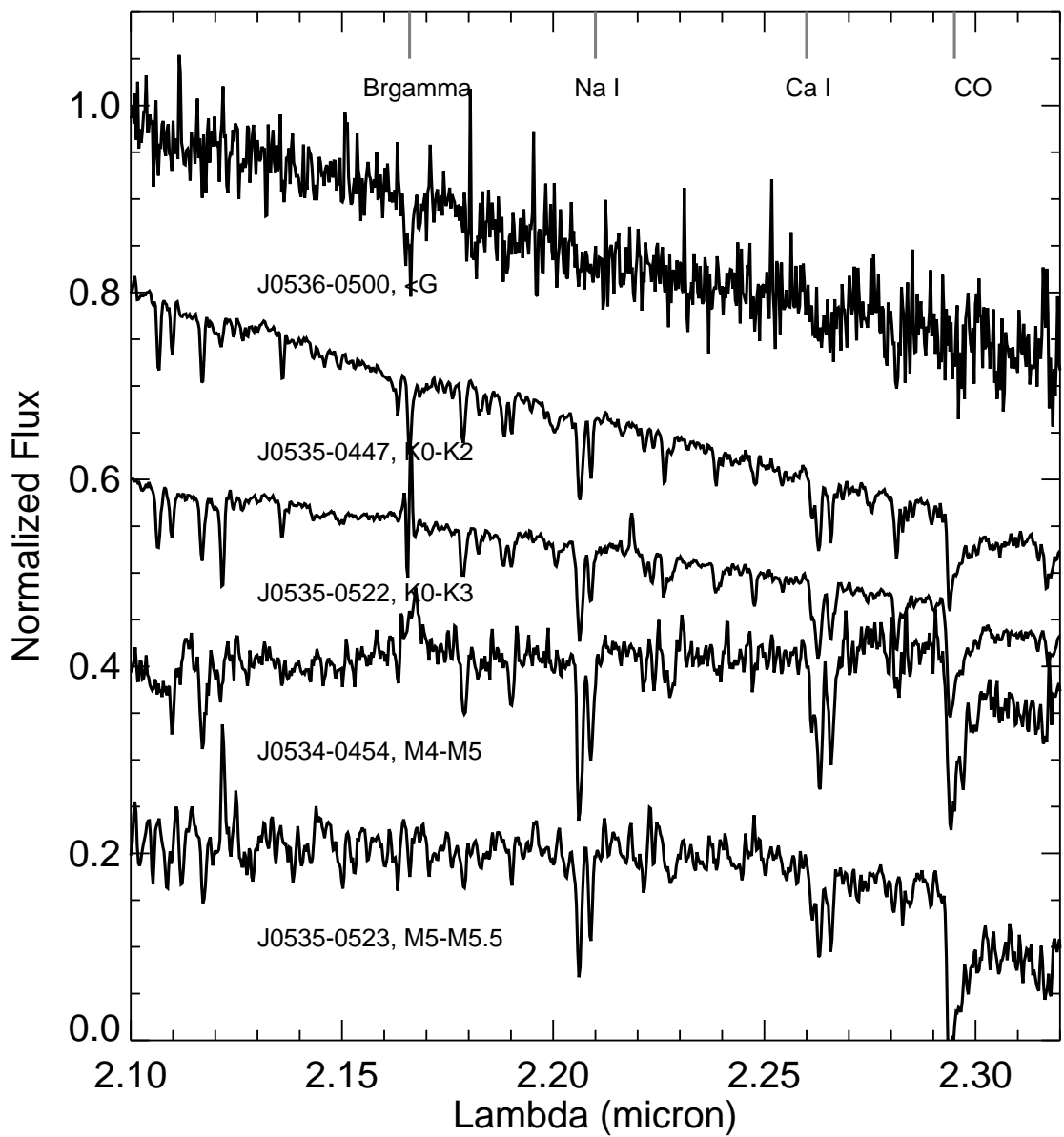}
\caption{TripleSpec K band spectra for ISOY J0536-0500, ISOY J0535-0447, ISOY J0535-0522, ISOY J0534-0454, and  ISOY J0535-0523 from top to bottom. The spectra have been normalized and offset in the y axis for clarity. The derived spectral types for each source are shown. The Bracket gamma and 2.12 micron H$_2$ lines in the spectrum of ISOY J0535-0522 and ISOY J0535-0523 respectively, are likely due to nebular emission. \label{TSpec}}
\end{figure}

We used these spectra to derive spectral types for our targets by comparison 
with dwarf standards in the in the Spex library \citep{Rayner09} and 
young TWA and Taurus members \citep{Luhman06b}. The derived spectral types are 
presented in Table~\ref{binaries} and the $K$ band region of the spectra are
shown in Figure~\ref{TSpec}.

\subsection{High-Resolution Spectroscopy}

On December 13, 2010 we obtained single epoch high resolution spectra for ISOY J0535-0447, ISOY J0535-0522, and ISOY J0535-0523, using the High Resolution Echelle Spectrometer \citep[HIRES;][]{Vogt94} at the Keck I telescope. In all cases we used the red cross-disperser and a 0.8$\arcsec$ width slit providing a resolution of R=$\lambda/\Delta\lambda=$50,000 and covering the wavelength range from 4310 to 8360 \AA. The exposure times were 120, 300, and 600 seconds respectively. In addition, for ISOY J0535-0447, we took an extra epoch on March 15, 2011 and on December 10, 2011 we took single epoch spectra for ISOY J0535-0525 and ISOY J0534-0449 with exposure times of 900 and 1800 respectively, keeping the same setup.

Data reduction was performed with ``MAKEE'' (MAuna Kea Echelle Extraction\footnote{http://www2.keck.hawaii.edu/inst/hires/makeewww/}), the standard data reduction package for the HIRES instrument which, starting from the raw images, produces wavelength calibrated spectra. A section of the spectra, showing the lithium absorption, for the 4 PMS EBs observed with HIRES can be seen in Figure~\ref{lithium}.

For ISOY J0535-0522, ISOY J0535-0525, and ISOY J0534-0454 we also obtained spectra with the NIRSPEC infrared echelle spectrograph \citep{McLean98,McLean00} at Keck II on November 9, 2011, January 11, 2012, and January 15, 2012 respectively. We used the $H$ band, centered at 1.555$\mu$m, with the 0.288$\times$24$''$ slit, yielding a resolution of $\sim$30,000. The total integration time was 8-16 minutes.
Data reduction was accomplished with the REDSPEC package\footnote{
http://www2.keck.hawaii.edu/inst/nirspec/redspec/index.html}.  Additional details are provided in \citet{Prato07}.

For ISOY J0535-0447, we have tried to obtain enough data to attempt to derive an orbit solution. For this purpose we obtained one additional high resolution spectrum using the echelle spectrograph at the Mayall 4m telescope at Kitt Peak National Observatory (KPNO) on September 24, 2010, and five epochs between November 22, 2010 and February 5, 2011 with NIRSPEC/Keck II .  Mayall 4m observations were obtained using the Echelle spectrograph with the `long red' camera, to cover 4540 - 7620 \AA  ~at an average resolution of 32,000.  
 We obtained two exposures of 900 seconds each. A standard echelle-data reduction was applied.
This includes background subtraction, cosmic-hit removal, flat-fielding and wavelength calibration. 
NIRSPEC observations were obtained using the setup described above and total integration times were 6 to 8 minutes.

We measured heliocentric radial velocities by cross correlation of the target spectrum with those of standard stars of similar spectral type using several orders. The accuracy of the radial velocities is 0.9-2 km/s, as measured by spectra of bright F and G stars of known radial velocity obtained on the same nights as the targets' spectra.  
The derived radial velocity measurements from each spectroscopic observation are listed in Table~\ref{spectable}. ISOY J0535-0525 and ISOY J0534-0454 showed clear double lines in their spectra. Both the H band and the optical spectrum of ISOY J0535-0522 are suggestive of a blend of two spectra but additional observations will be required to confirm the presence of double lines. None of the spectra of ISOY J0535-0447 showed double lines (see Sec. 5.2 for further discussion).

\section{Results and Discussion}

To put or EB systems into context we show in Fig~\ref{CMD} a $J$ vs. $J$-[3.6] color-magnitude diagram showing the location of the new PMS EBs (blue circles; $\theta^1$  Ori E is not shown due to the lack of shorter wavelength data) and the previously known ones (red squares) together with the Orion known members.

In this section we combine {\it Spitzer} light curve data with follow-up ground-based photometry and radial velocity
measurements, as well as data from the literature, to perform an initial
analysis of the physical parameters for each of the six PMS EBs discovered in 
the YSOVAR program. 
In cases where we have the most data, including well-sampled light 
curves at multiple wavelengths and radial velocity measurements, we 
attempt to measure the component stellar masses, radii, temperatures,
and other parameters. For most sources we have only the YSOVAR light 
curves, and data based on colors and/or spectral types, so our derived
properties are fairly rudimentary.  We present these 
preliminary analyses to guide follow-up investigations of these rare,
benchmark systems. For the remaining two EB systems, our analysis of the available data suggests that while eclipsing, the systems are not young members of Orion. We summarize the properties of these two systems in Appendix A.

\subsection{$\theta^1$ Ori E}

$\theta^1$~Ori~E, the ``fifth" member of the Orion Trapezium, is a known
double-lined spectroscopic binary  \citep[SB2;][]{Herbig06,Costero06,Costero08}. \citet{Costero08} found
that the two stars have nearly identical spectra, with spectral types of
G0~IV to G5~III (with a most likely spectral type of G2~IV). Strong 
\ion{Li}{1} $\lambda$6708 absorption and moderate \ion{Ca}{2}~K 
emission are observed, confirming the pre-main-sequence status
of the binary and, hence, its membership in the Orion Nebula
Cluster. Given the likely large stellar radii and relatively
tight orbit of the binary, it has been suspected that $\theta^1$~Ori~E 
might undergo eclipses \citep{Herbig06}. However,
its proximity to the brighter Trapezium star $\theta^1$~Ori~A has in
the past made it difficult to obtain reliable photometry.
The high spatial resolution and improved brightness contrast in the
infrared afforded by {\it Spitzer} makes an investigation of the
EB nature of $\theta^1$~Ori~E possible for the first time.

We show in Figures~\ref{PMSEBs_allLC2009} and \ref{PMSEBs_allLC2010} (middle right panels)
the {\it Spitzer} light curves obtained by us for $\theta^1$~Ori~E in 
2009 and 2010, respectively. Both light curves show periodic eclipses 
in both the [3.6] and [4.5] IRAC bands, 
showing for the first time that $\theta^1$~Ori~E
is indeed an EB. The eclipses are quite shallow,
indicating that this is a grazing eclipse.

In addition to the eclipses, there are indications of 
variations in the overall flux of the system. 
The median level of the {\it Spitzer} light curves is $\sim$64 mmag 
fainter in 2009 compared to 2010, whereas the [4.5] flux observed by us 
in 2010 is the same as was observed by Megeath et al. (2012) in 2004. 
We have explored the possibility that this small change in brightness 
is due to the photometry being affected by the bright nebulosity 
surrounding the Trapezium stars by checking the light curves of nearby 
sources. None of them showed a similar change in flux from one season 
to the next and thus we believe that the brightness change observed is real.
Indeed, $\theta^1$~Ori~E has also been
reported to exhibit intrinsic variation in previous studies at X-ray
and radio wavelengths.  \citet{Ku82} noted that the X-ray
flux fluctuated on a timescale of a few ks, and the nearly continuous
X-ray light curve spanning 13~d obtained in 2003 by the 
Chandra Orion Ultradeep Project \citep[COUP;][]{Stelzer05} showed a 
gradual 20\% brightening of the star followed by an abrupt spike
at about twice the original brightness. 
The star was also detected in Very Large Array observations at
2 and 6 cm by \citep{Felli93a, Felli93b}, during which variations of
$\sim$50\% in radio flux were seen. 
These X-ray and radio variations may not be surprising given the SB2 
nature of the system, however the cause of the infrared variations remains
unclear.   The most likely cause of these intrinsic flux variations
is stellar activity (star spots; flares; other coronal structures).
Here we simply note their existence and adjust the 2009 IRAC
light curve to match the continuum flux of the 2010 data in
order to permit an initial modeling of the infrared light curves. 
The time series data are provided in 
Table~\ref{EB45099_timeseries}. 

We combine these light curve data with the radial velocities measured by
\citet{Costero08} to produce a complete analysis of the binary. 
We used the \citeauthor{WD71} package (1971, updated 2005; hereafter WD)
as implemented in the PHOEBE code by \citet{Phoebe05}.
The code uses Kurucz stellar atmosphere models to represent the 
underlying stellar fluxes across the bandpasses, as well as the 
empirical limb darkening laws of \citet{vanHamme93}. The radial velocities
and light curve data are fit simultaneously and self-consistently.
The WD code does not include filter profiles or limb darkening 
coefficients for the IRAC bandpasses, so for this initial modeling, 
we approximated the [3.6] bandpass with the Johnson $L$ bandpass
and the [4.5] bandpass with the Johnson $M$ bandpass.
Strictly speaking, these differences between the bandpasses used for 
the observations and for the modeling introduce a systematic error 
in the model fluxes, but these differences should be small given the
approximate similarity and broadband nature of the bandpasses, as 
well as the survey-grade quality of the light curves used here.
The orbital period and time of periastron passage for the system were 
held fixed at the values previously determined by \citet{Costero08}.

\begin{figure}
\epsscale{.8}
\plotone{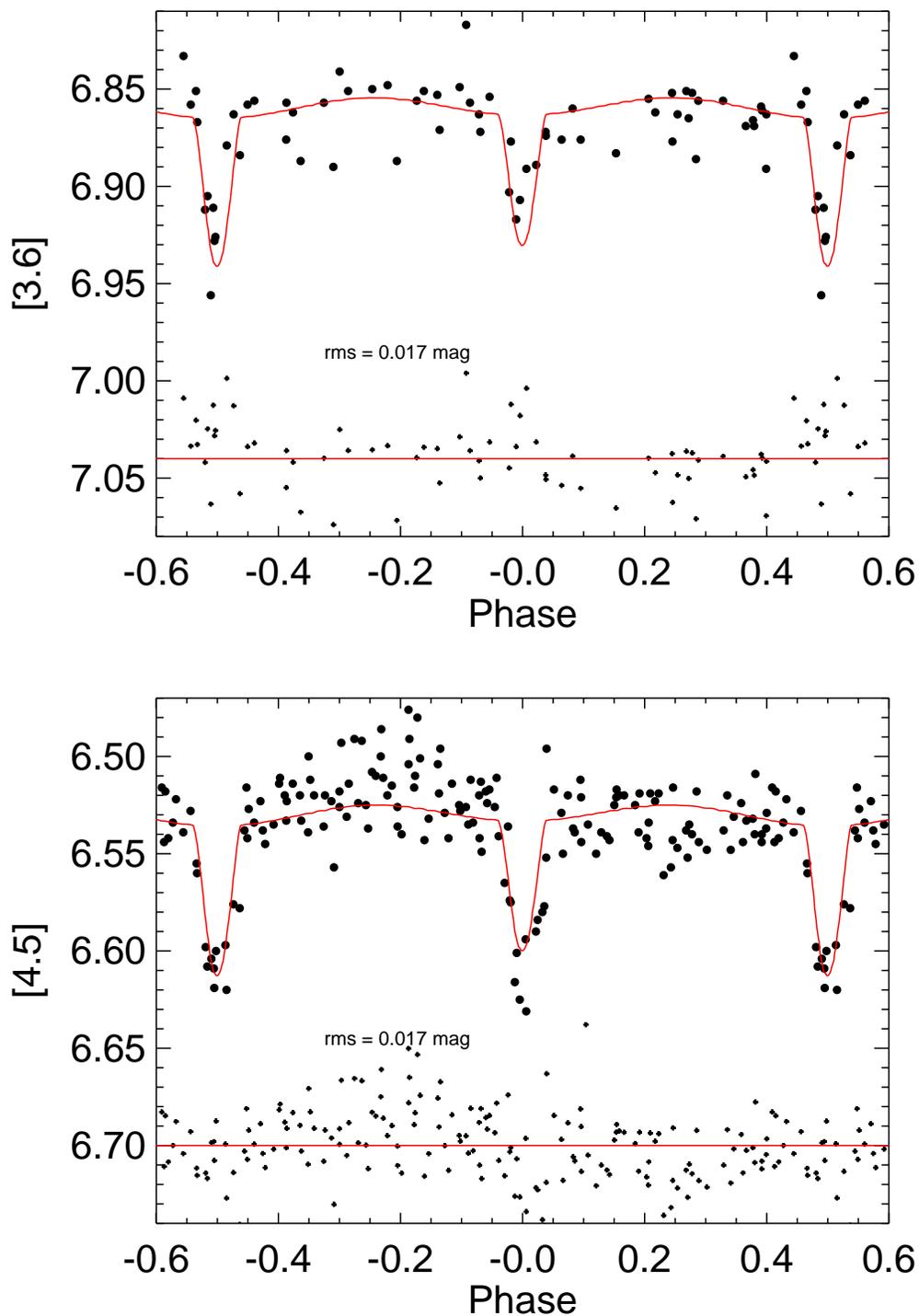}
\caption{Light curves for $\theta^1$ Ori E phased with a period of 9.89520 days in both IRAC channels, 3.6 $\mu$m (top) and 4.5 $\mu$m (bottom). The best fit model light curve is produced (see Sec 5.1) using the parameters from Table~\ref{ThetaOri_fit} and it is overplotted as a solid red line. The residuals of the fit can be seen in the lower part of each panel.\label{PhoebeThetaOri}}
\end{figure}

\begin{figure}
\epsscale{1.}
\plottwo{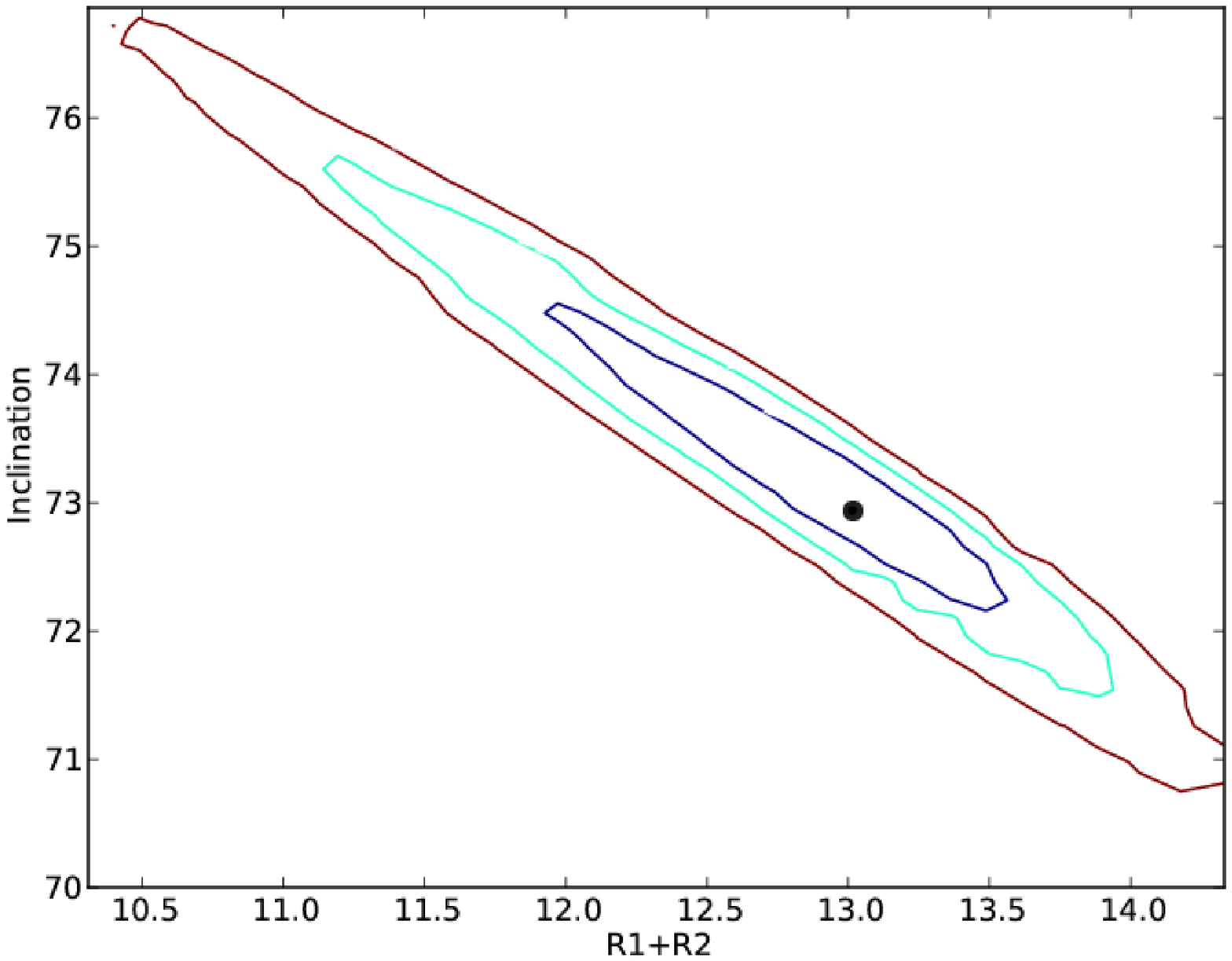}{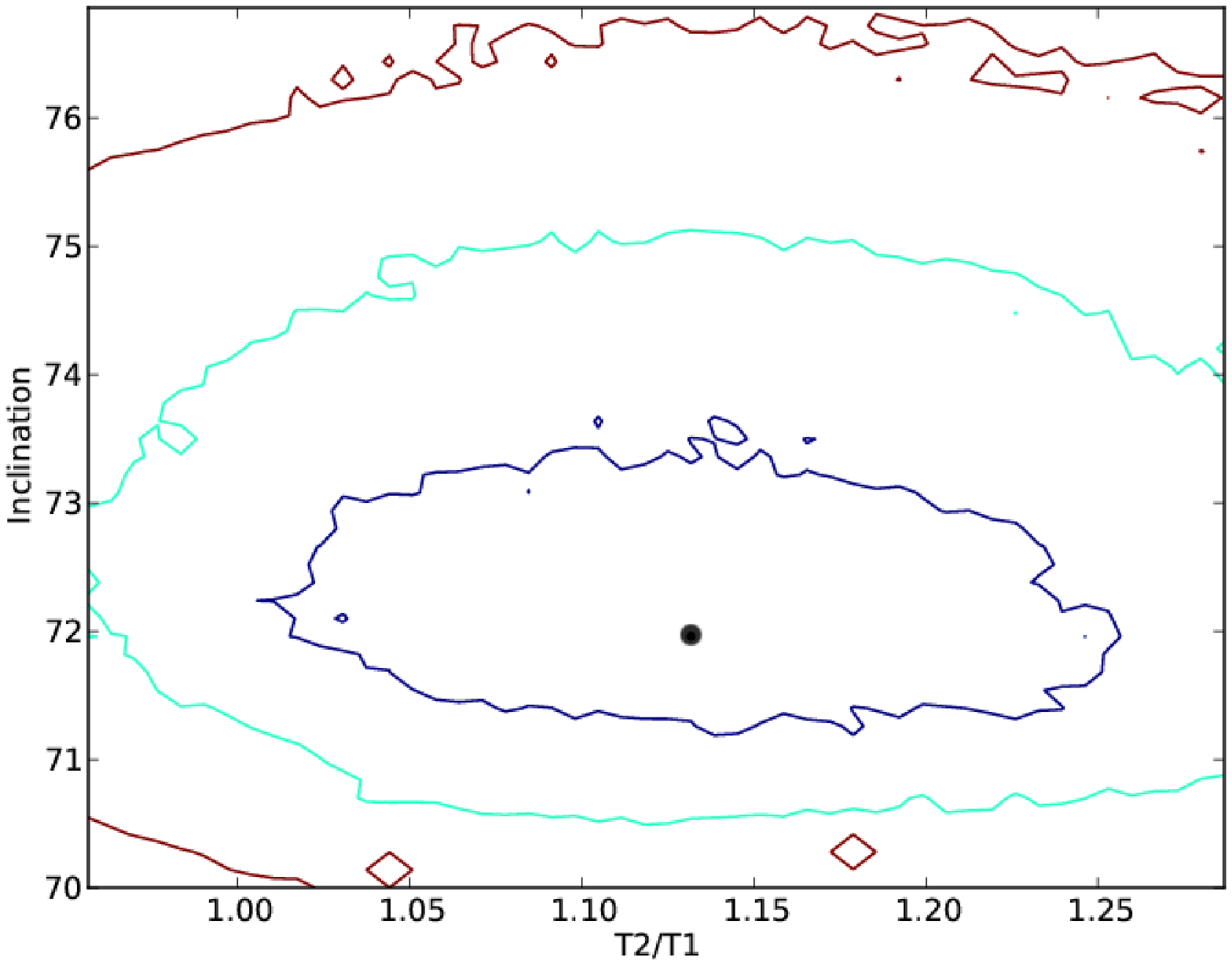}
\caption{Dependence of the radii (left) and  the temperature ratio (right) on the inclination for $\theta^1$ Ori E showing the degeneracy of the fitting process. The results of our fit are marked with a dot and the contours represent 1, 2, and 3 sigma confidence levels. Inclination and combined radius are highly degenerated; inclination and temperature ratio are less so, but also provide weaker constraints. \label{errorsthetaori}}
\end{figure}

The IRAC light curve fits resulting from our WD modeling are shown in
Figure~\ref{PhoebeThetaOri}, and the derived system parameters 
summarized in Table~\ref{ThetaOri_fit}.  
The precision and limited wavelength coverage of our light curve data 
do not permit highly precise determination of the stellar radii or 
temperatures, however we are able to provide constraints on the sum 
of the stellar radii and on the ratio of the components' effective temperatures.
These parameters and the system inclination show some degeneracy
(Figure~\ref{errorsthetaori}), which we incorporate into
our parameter uncertainties in Table~\ref{ThetaOri_fit}.
The mass ratio of the system is extremely well determined from the 
excellent quality radial velocity data of \citet{Costero08}, for
which we find $q=0.9965\pm0.0065$; the two stars are of very nearly
identical mass.
The system inclination is now also very well constrained from our light 
curve fits (i$\sim$74$^\circ$), 
and thus we are able to report individual masses for $\theta^1$~Ori~E of
$M_1$=2.807 M$_\odot$ and $M_2$=2.797 M$_\odot$, accurate to $\sim$2\%.
It is thus the highest mass EB yet discovered with clear PMS nature.

The temperature ratio that we determine for $\theta^1$~Ori~E is T$_2$/T$_1$=1.12 $\pm$ 0.08.
This is roughly consistent with the expectation of identical temperatures
if the stars are of truly identical mass; however, the measured mass
ratio also allows for a small mass difference which would then also
accommodate a small temperature difference. Higher precision light
curves over a larger range of wavelengths will be required to more
firmly pin down the temperature ratio. The inferred temperature ratio 
is also modestly dependent upon the adopted absolute temperatures of
the stars. Here we have assumed $T_1 = 6000$~K based on the reported
spectral type, but this too needs to be improved.

The sum of the radii that we measure is $R_1+R_2=12.5 \pm 0.6$ R$_\odot$. 
It is premature to attempt a detailed comparison to the predictions
of pre--main-sequence stellar evolution models, however these radii
appear to be broadly consistent with expectations.  For example,
the model isochrones of \citet{Siess00} predict a radius sum of
11.1--14.5~R$_\odot$ for ages in the range 0.3--2~Myr.

\subsection{ISOY J0535-0447} 

ISOY J0535-0447 is a known proper-motion member of the ONC 
\citep{Tian96}, and our own spectroscopy shows strong lithium absorption
(see Figure~\ref{lithium}),
further confirming ISOY J0535-0447 as a PMS member of the ONC. From our TripleSpec data (see
Figure~\ref{TSpec}) we derive a spectral type of K0-K2 and extinction of $A_v$=1.5-2 based on a comparison to SpeX dwarfs \citep{Rayner89}. Our HIRES spectrum yields a spectral type of K1 ($\pm$ 1 subtype) based on comparison with dwarf standards.
This star has also been included in previous variability surveys of 
the ONC, but has not heretofore been identified as an EB.  
Neither the near-IR variability survey of \citet{Carpenter01}
nor the Chandra X-ray variability survey of \citet{Ramirez04}
found ISOY J0535-0447 to be photometrically variable. 
Similarly, \citet{Tobin09} did not flag this star 
in their radial-velocity survey for SBs in the ONC. 

\begin{figure}
\epsscale{1.}
\plotone{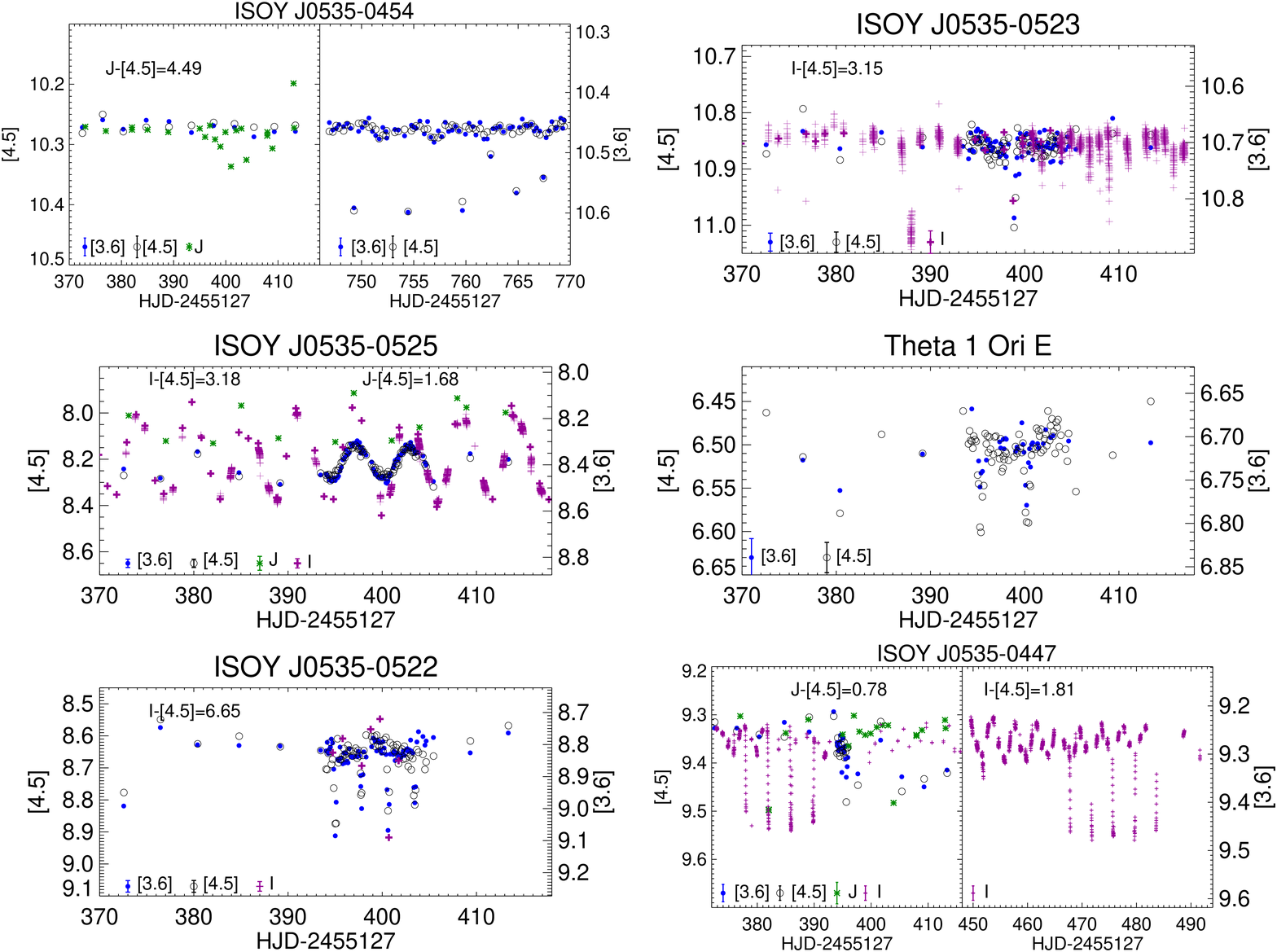}
\caption{Light curves for the 6 new Orion PMS EBs discovered by the YSOVAR program from the follow up observations in 2010 and 2011. The symbols are $\bullet$: [3.6]; $\circ$: [4.5]; $*$: $J$ (UKIRT); and $+$: $I_c$ (APO). [3.6] and [4.5] magnitudes have been plotted in the right and left vertical axis respectively. $I_c$ and $J$ light curves, if present, have been shifted in the y-axis to match the mean IRAC values. Mean colors are stated in each panel. Note that the x-axis is different depending on the available data.\label{PMSEBs_allLC2010}}
\end{figure}

That the EB nature of this star was previously missed both photometrically
and spectroscopically is likely due to a confluence of several factors:
the orbital period is very close to an integer number of days
($P\approx3.9$~d; see below), the optical eclipses are relatively shallow 
(see Figure~\ref{PMSEBs_allLC2010}, lower right panel), 
the secondary eclipse is extremely shallow and only readily detectable
in the IRAC bands,
and the optical spectrum does not reveal a secondary spectrum but
instead appears as a single star of spectral type $\sim$K1. 

The secondary eclipses in this system are sufficiently shallow that
from our discovery Spitzer light curves it was not immediately evident 
whether secondary eclipses were present at all. In addition, our intensive
follow-up light curves from the ground in $I_c$-band, which are of excellent
precision, exhibit out-of-eclipse variations whose amplitude is larger
than the secondary eclipse and which vary with a period slightly different
from the orbital period (see below), making ready identification of the 
secondary eclipse difficult. Therefore we obtained a follow-up IRAC
light curve at the expected secondary eclipse time
(see Figure~\ref{EB3227_allLC2010})
definitively proving the existence of a faint secondary star, and 
providing a very strong 
constraint for the modeling of the system, which we now describe.

The most comprehensive light curve data were obtained with the USNO 40$\arcsec$ telescope at 
$I_c$ band, with high cadence sampling over several hours per night on
about 70 nights in 2010 and 2011. 
Given the strong out-of-eclipse variations present in the $I_c$-band
data which might dominate over the eclipse signal with Fourier based
period-search methods, we determined the eclipse period 
using a BLS period-finding algorithm and then we manually adjusted
the resultant best period by combining the data from the different 
bandpasses and adding the monitoring data from \citet{Rebull01}. We determine the following system ephemeris, which we
adopt throughout our analysis:
\begin{center}
HJD$_0$=2455126.26  \\
P=3.905625$\pm$0.000030 days\\
\end{center}
\noindent

As mentioned, the $I_c$ band light curve exhibits a strong 
periodic modulation superposed on the eclipses. In addition, 
a flare is visible around HJD 2455502 (day 375 in Fig.~\ref{PMSEBs_allLC2010}).
These features strongly indicate the presence of spots on the primary 
star modulated on its rotation period (the secondary is very
likely too faint to produce such a strong spot modulation signal). 
We find this spot modulation to have a period slightly longer than 
the EB period (4.001~d vs.\ 3.906~d), indicating a modest departure
from full synchronization of the system, and which produces the
stroboscopic effect that causes the apparent secular changes in the
primary eclipse depths over time (Figure~\ref{PMSEBs_allLC2010}).
In addition, we found that the spot modulation is not well fit by
a single sinusoid, but rather appears to have two sinusoidal 
components that moreover evolve modestly in their relative amplitudes
and phases.  This is
presumably due to changes in the spot coverage of the stellar surface
causing shape and amplitude changes in the out-of-eclipse modulations.

\begin{figure}
\epsscale{.45}
\plotone{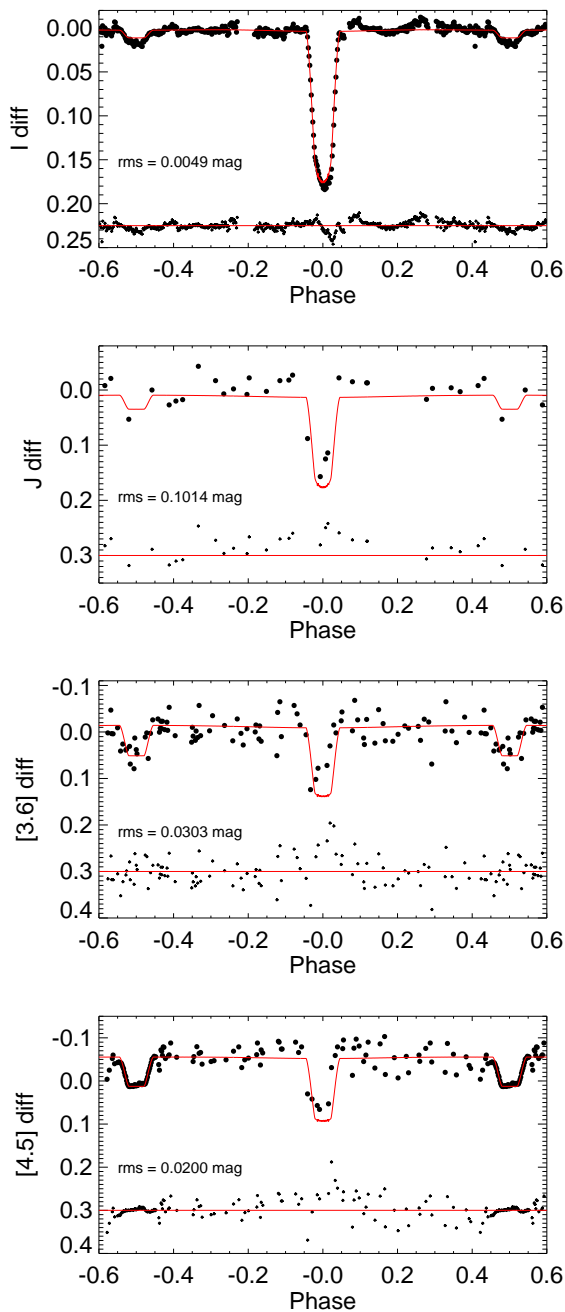}
\caption{Light curves for ISOY J0535-0447 folded with a period of 3.905625 days. From top to bottom: $I_c$-band (rectified as described in Sec. 5.2), $J$-band, 3.6 $\mu$m, and 4.5 $\mu$m. The best fit model light curves are produced for each filter using the parameters from Table~\ref{EB3227_fit} and are overplotted as solid red lines. Note that the IRAC [4.5] staring data around the secondary eclipse provides a very strong constraint for the modeling of the system. The residuals of the fit can be seen in the lower part of each panel. Even around the secondary eclipse the fit is not perfect due to the stellar variability. \label{PhoebeEB3227}}
\end{figure}

In principle it should be possible to model the system light curves
incorporating a full physical spot model. For this initial investigation,
we have performed a rectification of the $I_c$-band light curve in an
attempt to remove the complex out-of-eclipse variations. 
We fit the light curve with a two-component Fourier model, with a 
single principal period of 4.001~d (see above).
To allow for the possibility that the
variability changed in amplitude and/or phase on long timescales, we
fit and subtracted the model to the 2010 and 2011 $I_c$-band data 
separately. 
Overall, the light curve fits are reasonably good (see
Figure~\ref{PhoebeEB3227}). Some unmodelled variation in the $I_c$-band 
light
curve still remains and the residuals affect the primary eclipse slightly
and perhaps also the secondary eclipse depth. The model for the primary eclipse at longer wavelengths in Figure ~\ref{PhoebeEB3227} appears by eye to overestimate the eclipse depth.   Due to the uncertainties and limited sampling in the photometry for the primary eclipse at these wavelengths, relative to the higher-precision, higher-cadence staring mode data of the secondary eclipse, our fitting routine places more weight on the accuracy of the fit to the secondary eclipse.  The primary eclipse is well fit in our shortest wavelength, and thus it is unlikely that there are large systematic errors in the secondary and/or primary radii.  However, there may be some small errors in the assumed extinction-corrected colors of the components in this system.

None of the spectra taken for this EB candidate show double lines; 
however, they do show small radial velocity (RV) variations 
(see Table~\ref{spectable}).  The small RV amplitude together with the
very shallow secondary eclipse depth
point to the companion being much lower mass than the primary. 

\begin{figure}
\epsscale{.6}
\plotone{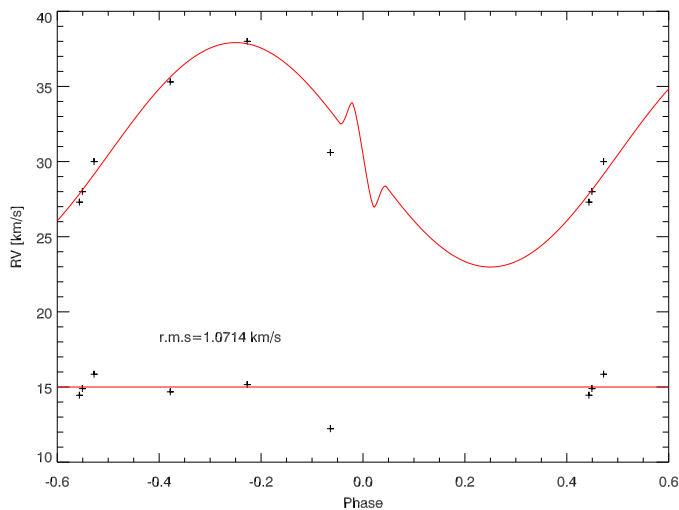}
\caption{Radial velocity measurements for the primary star of ISOY J0535-0447. A model radial velocity curve is generated using the parameters from Table~\ref{EB3227_fit} and is overplotted with a solid red line. The residuals of the fit are shown in the bottom of the panel and have an rms of 1.07 km/s. \label{PhoebeEB3227rvs}}
\end{figure}

We have used the few available radial velocity measurements,
together with the light curve data, for a first attempt to obtain the
physical parameters of the system.  The photometric times of minima
indicated no sign of orbital eccentricity, which is expected for
a short-period binary such as this, so we assumed circular orbits in the
light curve analysis. In addition, since we only have RV measurements for the
primary, we had to make one more assumption for the orbital solution. Thus,
we adopted a total mass of $\sim$0.9 M$_\odot$ (by fixing the semi-major axis, a=10R$_\odot$) and then fit for
the mass ratio and the center-of-mass velocity. In Table~\ref{EB3227_fit}
we present the orbital parameters of ISOY J0535-0447 resulting from our
best orbit solution and display this solution along with the radial-velocity
measurements in Figures~\ref{PhoebeEB3227} and~\ref{PhoebeEB3227rvs}. Note
that our derived center-of-mass velocity for ISOY J0535-0447 of $\gamma$
= 30.4 km/s is slightly larger than the currently accepted ONC systemic
cluster velocity of 25$\pm$ 2 km/s \citep{Sicilia05}. The presence of
spots on the stellar surfaces can also introduce distortions that affect
the observed radial velocities; however, \citet{Stassun04} found that the
spot effect on the RVs is $\sim$1 km/s and probably negligible for our data.

With estimated masses for the system
components of 0.83 M$_\sun$ and 0.05 M$_\sun$, the secondary seems to be in the brown dwarf regime, filling an
important gap in the current census of PMS EBs. Note however that this is a
preliminary result; the inferred temperature of the secondary is slightly warmer than
what would be expected for a brown dwarf. We have an ongoing study of this
system to further characterize the physical properties of the components
of ISOY J0535-0447 and measure their masses more accurately.

\subsection{ISOY J0535-0522}

ISOY J0535-0522 is located at about 1$\arcmin$ northeast of the Trapezium. It has been included in several previous studies, but it is in a very complicated region with bright structured nebulosity which has probably prevented its identification as an EB before. 
ISOY J0535-0522 was detected as an X-ray source and hence confirmed as a 
probable member of the ONC by COUP \citep{Getman05}.  It was also included in the 
near-IR variability study performed by \citet{Carpenter01} where it was 
not found to be variable, however, it was flagged as a possible flaring star 
by \citet{Feigelson02} in their study of X-ray-emitting young stars in the Orion Nebula.

\begin{figure}
\epsscale{1.}
\plotone{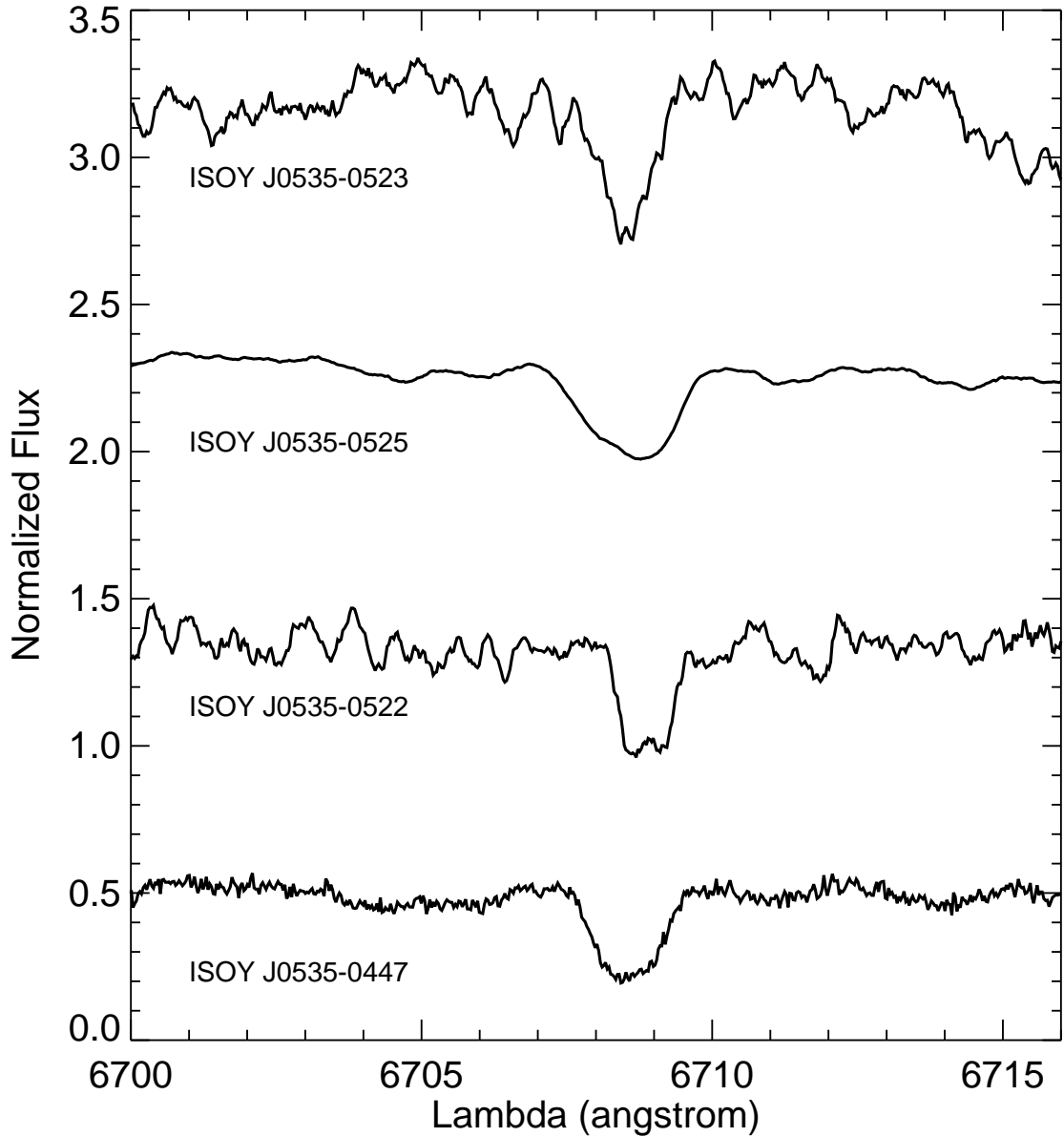}
\caption{HIRES spectra for the four PMS EB for which we have high resolution optical spectra showing the presence of lithium absorption. From top to bottom: J0535-0523, J0535-0525,  J0535-0522, and  J0535-0447. The spectra have been normalized and offset in the $y$ axis for clarity. \label{lithium}}
\end{figure}

ISOY J0535-0522 was classified as a late K or early M star by \citet{Luhman00}, 
however, our TSpec data (see $K$-band in Figure~\ref{TSpec}) is very similar to 
that of ISOY J0535-0447, but with higher extinction (A$_v$=11.5-12), suggesting a
spectral type near K0. The hydrogen lines appear contaminated by nebular emission and without 
those lines a K3 spectral type cannot be ruled out. A spectral type of K0-K3 and 
large $A_v$ are more consistent
with the location of the system in a color-magnitude diagram (see Figure~\ref{CMD}) 
than the previously reported spectral type.
A Keck/HIRES single epoch spectrum yields a spectral type of K0 $\pm$ 2 subtypes and shows CaII core emission in the triplet lines and lithium absorption (see Figure~\ref{lithium}), both features confirming 
its PMS status. Both the NIRSPEC and HIRES spectra of ISOY J0535-0522 are suggestive of the SB2 nature of the system (see the HIRES spectrum in Figure~\ref{lithium}) and the IRAC 
[3.6] and [4.5], and $I_c$ band light curves from 2010 confirm the eclipses discovered in 2009. 

\begin{figure}
\epsscale{1.}
\plotone{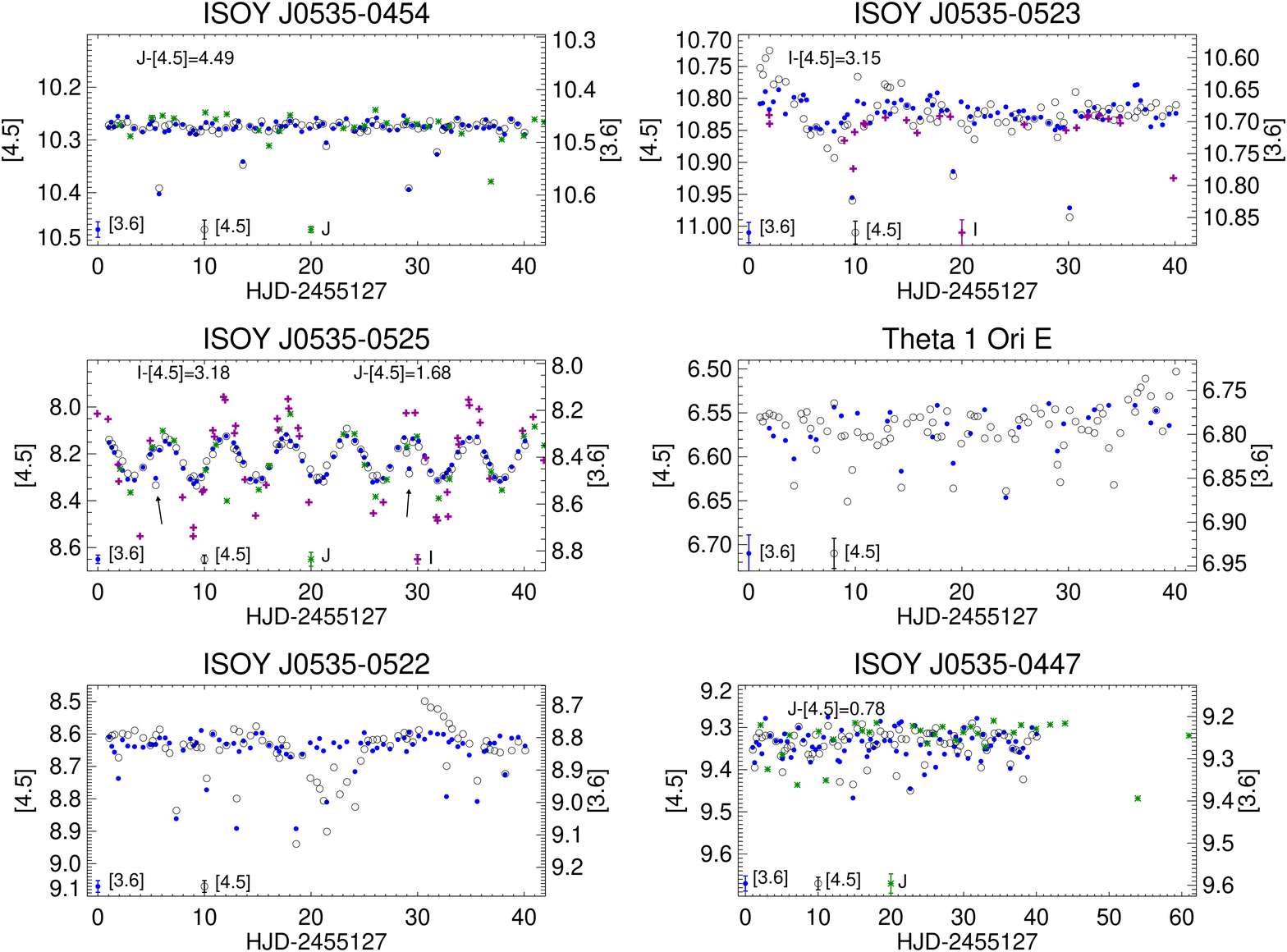}
\caption{Light curves for the 6 new Orion PMS EBs discovered by the YSOVAR program from 2009 data. The symbols are $\bullet$: [3.6]; $\circ$: [4.5]; $*$: $J$ (UKIRT); and $+$: $I_c$ (APO \& LOWELL). [3.6] and [4.5] magnitudes have been plotted in the right and left vertical axis respectively. $I_c$ and $J$ light curves, if present, have been shifted in the y-axis to match the mean IRAC values. Mean colors are stated in each panel. Note that the x-axis is different depending on the available data. \label{PMSEBs_allLC2009}}
\end{figure}

The light curves of this EB candidate are presented in 
Figures~\ref{PMSEBs_allLC2009} and~\ref{PMSEBs_allLC2010} (lower left panels) for the 2009 and 2010 
observing runs respectively and the time series are available in Table~\ref{EB46222_timeseriesI1}. The 2009 light curves show a trend in the 
[4.5] data between HJD 2455147 and 2455152 (between days 20 and 25 in Fig.~\ref{PMSEBs_allLC2009}) and between HJD 2455157 and 2455161 (between days 30 and 34 in Fig.~\ref{PMSEBs_allLC2010}) that is not followed by the [3.6] data and we believe is not real. 
Those datapoints were excluded from the following analysis. The 2010 data did not 
show similar artifacts but did show an ascending trend that lasted for all the 
2010 period. Again, the [3.6] data did not show the same behavior and, given that 
the nebulosity is brightest at [4.5], we believe that our photometry was affected 
by it. We corrected the 2010 IRAC [4.5] time series by subtracting a linear fit 
to the whole raw light curve in 2010 (the light curve shown in 
Figure~\ref{PMSEBs_allLC2010} as well as the time series provided in Table~\ref{EB46222_timeseriesI1} are already corrected from that effect).  

The light curves include over 10 clearly discernible eclipses  which have 
allowed us to perform a periodogram analysis using the BLS period-finding 
algorithm. The resulting ephemeris is:
\begin{center}
HJD$_0$=2455128.79 \\
P=5.6175$\pm$ 0.0015 days\\
\end{center}

\begin{figure}
\epsscale{.6}
\plotone{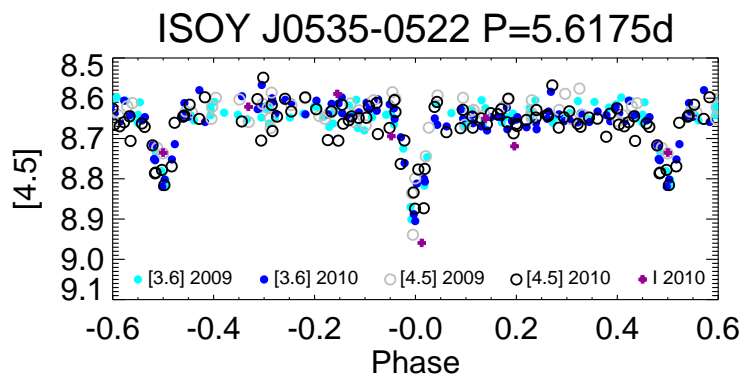}
\caption{Phased light curve for ISOY J0535-0522 folded with a period of 5.6175 days where all the data from 2009 and 2010 are plotted. The symbols are $\bullet$: [3.6] IRAC; $\circ$: [4.5] IRAC;  and $+$: $I_c$ WFI;. [3.6] and $I_c$ light curves have been shifted in the y-axis to match the mean [4.5] values ([3.6]-[4.5]=0.31 mag, $I_c$-[4.5]=6.65 mag). \label{46222phased}}
\end{figure}

Using these parameters, we present the folded light curves in Figure~\ref{46222phased}. 
A rough comparison of the primary and secondary eclipse depth yields a 
temperature ratio of T$_2$/T$_1 \sim$ 0.94 which, assuming a temperature 
of 5250 K for the K0 primary \citep[based on the PMS temperature scale of ][]{Kenyon95}, produces a temperature of $\sim$4935 K for the secondary, which would then be a K2 star.

\subsection{ISOY J0535-0523}

ISOY J0535-0523 was labeled as a proper motion member of the ONC by \citet{Jones88} and it is 
located close to the Trapezium stars as well ($<$3$\arcmin$ to the west of the Trapezium). 
It has been labeled as non-variable by both \citet{Carpenter01} and \citet{Feigelson02}. 
In 1994, \citet{Stassun99} performed an $I_c$ band monitoring of the ONC;
a similar study was performed by the 
Monitor Project during 2004 to 2006 \citep{Irwin07} - neither group identified the star as being an EB. 

A spectral type of M4.5 has been previously reported \citep{H97} and our TSpec 
data (see Figure~\ref{TSpec}) show that it is indeed a mid-M star, M5-M5.5, with 
A$_v$=1.5 based on a comparison of steam bands to members of TWA and Taurus. This is in agreement with the location of ISOY J0535-0523 in the color-magnitude diagram showed in Figure~\ref{CMD}. Our 
single epoch HIRES spectrum yields an spectral type of M6 $\pm$ 1 subtype and shows lithium absorption as can be seen in Figure~\ref{lithium}. The spectrum does not show obvious double lines.

The light curves for 2009 and 2010 are shown in Figures~\ref{PMSEBs_allLC2009} 
and~\ref{PMSEBs_allLC2010} where a handful of  eclipses are noticeable. We have 
added the dataset from \citet{Stassun99} and the Monitor Project to our $I_c$ band 
observations in order to use the long time baseline to improve the accuracy of 
our period analysis. From these light curves we have derived a tentative ephemeris as:
\begin{center}
HJD$_0$=2455136.7\\
Period=20.485$\pm$ 0.5days.\\
\end{center}

\begin{figure}
\epsscale{.6}
\plotone{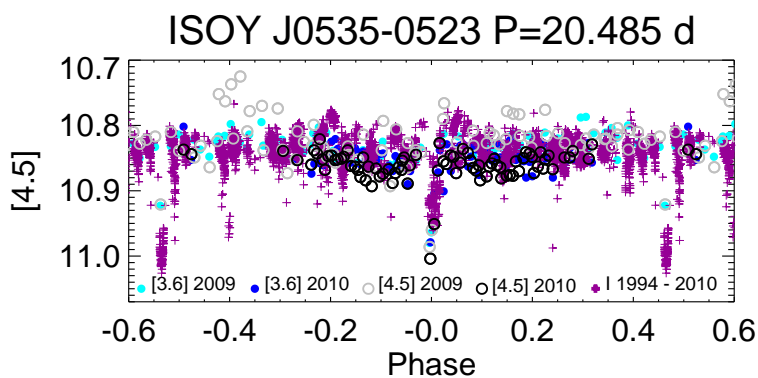}
\caption{Phased light curve for ISOY J0535-0523 folded with a period of 20.485 days where all the data from 2009 and 2010 are plotted. The symbols are $\bullet$: [3.6] IRAC; $\circ$: [4.5] IRAC;  and $+$: $I_c$ APO+WFI+\citet{Stassun99}+ the Monitor Project. [3.6] and $I_c$ light curve have been shifted in the y-axis to match the mean [4.5] value ([3.6]-[4.5]=0.05, $I_c$-[4.5]=3.15). Note that there are some $I_c$ band data around phase=-0.4 that doesn't line up. Our photometry phases well with the Monitor Project data but the data from \citet{Stassun99} shows some disagreement probably due to noisy photometry or a small error in the period accumulated over the years (see Sec. 5.4)\label{40134phased}}
\end{figure}

\noindent However, among all the data, only one eclipse was well mapped. This and the fact that the eccentricity is not exactly zero makes it more difficult to pin down the period. Thus, we provide above the most likely period and note that the error in the period determination for this source is larger than for the other systems. The folded light curve is provided in Figure~\ref{40134phased}.  The Monitor Project $I_c$ band data (from 2004) confirm the eclipse signatures seen in our data, however there is some disagreement with the $I_c$-band data from \citet{Stassun99} (at phase$\sim$-0.4) probably due to a small error in the period accumulated over the years. Due to the scattering of the photometry in the eclipses, it is not straight forward to determine which one is the primary or the secondary and thus an estimation of the temperature ratio is not possible.

ISOY J0535-0523 has a period far longer than any of the known PMS EBs and much 
longer than the expected rotational period. It may thus be the only known EB that is 
probably not spun up by tidal effects. If dynamo activity increases the radii of 
most EB stars, ISOY 0535-0523 could be the only system known whose components' radii are not inflated by these tidal effects, thus 
allowing its properties to be better compared to the current generation of theoretical 
models providing a direct test of the link between magnetic fields, 
rotation, and radii. Performing this test will require an RV solution in order to fully characterize the system. 

\subsection{ISOY J0534-0454 }

\begin{figure}
\epsscale{.8}
\plotone{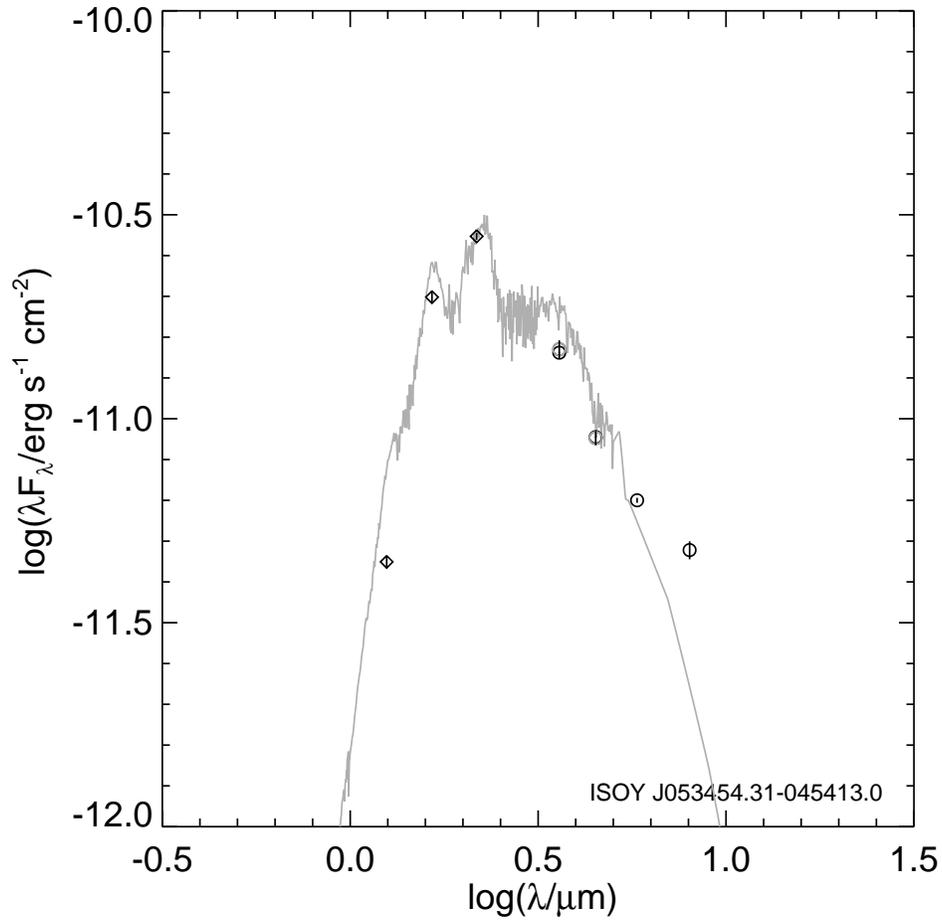}
\caption{Spectral energy distribution for ISOY J0534-0454 showing the presence of a moderate mid-IR excess indicative of a circumstellar disk. The combined photometry for the system is used. The data come from 2MASS (diamonds) and IRAC (black circles for data coming from Megeath et al. 2012 and grey circles being the median values from our YSOVAR light curves). For comparison a reddened M5 Kurucz model normalized at K band is shown.\label{1512SED}}
\end{figure}

ISOY J0534-0454 is catalogued as having a disk based on its mid-IR excess 
(Megeath et al. 2012). Its spectral energy distribution (SED) can be seen in 
Figure~\ref{1512SED}. ISOY J0534-0454 is the only EB candidate that we found 
among the disked objects. It has been labeled as non variable both by \citet{Carpenter01} and \citet{Ramirez04}. 

ISOY J0534-0454 is heavily extincted and so we lack optical photometry. Our TSpec data indicates a mid-M spectral type, which is consistent with its location on a $J$ vs. $J$-[3.6] color-magnitude diagram (see Figure~\ref{CMD}) assuming high extinction. The Keck/NIRSPEC data shows double lines in the spectrum indicating that the system is an SB2.
The light curve of ISOY J0534-0454 in 2009 (see Figure~\ref{PMSEBs_allLC2009}) 
shows a handful of eclipse features clearly seen at least at IRAC wavelengths 
(which has a higher cadence than the $J$ UKIRT data). For this system, we do not 
have high cadence observations in 2010 (Fig.~\ref{PMSEBs_allLC2010}) and the 
lower cadence data ($\sim$12 datapoints in 40 days) does not show any indication of the presence of an eclipse. However the Spitzer data from 2011, where a few eclipses are present, further confirm the EB nature of the system and yield the ephemeris as:
\begin{center}
HJD$_0$=2555127.56\\
Period=5.1993$\pm$ 0.0004 days.\\
\end{center}

\begin{figure}
\epsscale{.6}
\plotone{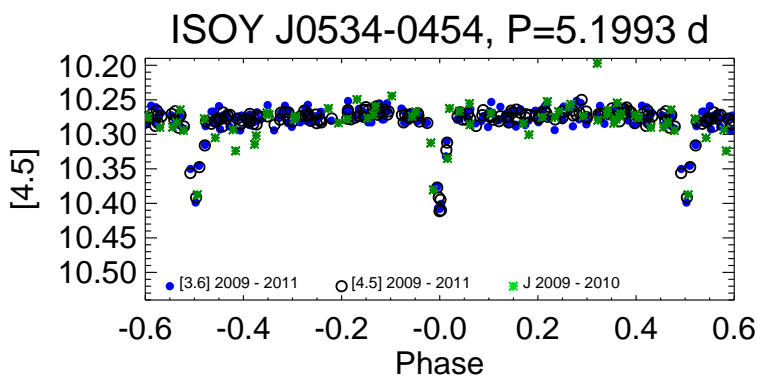}
\caption{Phased light curves for ISOY J0534-0454 folded with a period of 5.1995 days where all the data from 2009 to 2011 are plotted. The symbols are $\bullet$: [3.6] IRAC; $\circ$: [4.5] IRAC;  and $*$: $J$ UKIRT. [3.6] and $J$ light curves have been shifted in the y-axis to match the mean [4.5] value ([3.6]-[4.5]=0.21, $J$-[4.5]=4.49).  \label{1512phased}}
\end{figure}

\noindent The folded light curve using that period can be seen in 
Figure~\ref{1512phased}. We note that the light curve is remarkably clean despite the presence of a disk. ISOY J0534-0454 seems to be a system with $e\sim$0 and with two similar components with T$_2$/T$_1 \sim$ 0.96 based on the eclipse depth. Assuming a temperature of 3125~K for an M5 primary using the \citet{Luhman99} temperature scale for young M stars
yields a temperature of $\sim$2990~~K for the secondary, and thus an M6 star. The secondary may then be a brown dwarf given its spectral type (the known brown dwarf EB in Orion is an M6.5).

\subsection{ISOY J0535-0525}

ISOY J0535-0525 was selected as a proper motion member of the ONC by \citet{Jones88}.  
It has been labeled a long term variable by \citet{Feigelson02}; \citet{Herbst00} 
found that it was a periodic variable with a period of $\sim$5.7 days, which
is consistent with our photometry. A wide range of spectral types are assigned to this target in the literature:  M0.4, K4, and K7-M1 \citep{H97}, probably due to its location close to the Trapezium cluster and the variable extinction in the region. It was included on the radial-velocity survey for SBs in the ONC done by  
\citet{Tobin09} where it was not flagged as a probable binary (V$_{rad}$=22.9$\pm$2.25). However,  our high resolution spectra of this source shows the presence of double lines and thus confirms its binary nature. Furthermore, the HIRES
spectrum we obtained shows it to have a strong lithium absorption feature,
confirming its youth (see Figure~\ref{lithium}). Thus, we can classify it as a PMS EB.

However, only our 2009 IRAC data show
evidence for this designation (see Figures~\ref{PMSEBs_allLC2009} and~\ref{PMSEBs_allLC2010}).   The 2009 IRAC light curve
shows a well-defined approximately sinusoidal variation presumably reflective of
rotational modulation of a spotted star.  However, at two epochs - HJD 2455132 and 2455156 (days 5 and 29 in Fig.\ref{PMSEBs_allLC2009}) -
the photometry of both IRAC channels show a 0.15 mag dip of duration less than
one day.  We do not have shorter-wavelength data at the times of these dips.
If the dips are due to eclipses, and if no eclipses were missed in the IRAC data,
then the period would be of order 24 days.  There is another event at HJD 2455528 (day 401 in Fig.~\ref{PMSEBs_allLC2010}) and several other single point deviants that we cannot confirm as eclipse features.

This could be another very useful PMS EB if its period were indeed confirmed
to be $\sim$24 days, since it would then be the longest period PMS EB known and
could serve as an additional empirical test of PMS isochrones for objects
that are not tidally locked to short rotation periods.

\section{Conclusions}

We have identified six new EB star systems that are
believed to be members of the Orion Nebula Cluster and thus are PMS EBs.  For one of them, $\theta^1$ Ori E, we provide an orbital solution with a complete set of well determined parameters. The masses derived for its components are 2.807 and 2.797 M$_\sun$ and thus $\theta^1$ Ori E is the most massive EB known to date with clear PMS nature. For a second system, ISOY J0535-0447, we provide a preliminary orbital solution based on our light curves and
available radial velocity data. For the other four, we provide
periods and relatively well-defined light curves, often based on multi-year
data.  One of these systems, ISOY J0535-0523, is the longest period (P$\sim$20 days) PMS EB currently known. Such a long period suggests that the components will be less affected by tidal effects providing a direct test between magnetic fields, rotation and radii. Another two systems seem to have secondary components in the brown dwarf domain providing, if confirmed, additional examples to the only one known PMS brown dwarf EB.

These systems increase the number of known PMS EBs by over 50\%, and the unique properties of several of these systems ensure that they will offer the potential for considerably improving the empirical calibration of the PMS models for low mass stars and brown dwarfs. However, additional observations are needed to fully characterize the systems, in particular high resolution spectroscopy at several epochs is needed to derive the orbital parameters of these systems.




\acknowledgments

We are grateful to the anonymous referee for a helpful report. This work is based in part on observations made with the Spitzer Space Telescope, which is operated by the Jet Propulsion Laboratory, California Institute of Technology under a contract with NASA. Support for this work was provided by NASA through an award issued by JPL/Caltech.
This research has made use of the NASA Exoplanet Archive, which is operated by the California Institute of Technology, under contract with the National Aeronautics and Space Administration under the Exoplanet Exploration Program.
K. L. was supported by grant AST-0544588 and L. P. by grant AST-1009136 from the
National Science Foundation.
The Center for Exoplanets and Habitable Worlds is supported by the
Pennsylvania State University, the Eberly College of Science, and the
Pennsylvania Space Grant Consortium. 
S. M. was supported by the Gemini Observatory, which is operated by the Association
of Universities 
for Research in Astronomy, Inc., on behalf of the international Gemini
partnership of Argentina,
Australia, Brazil, Canada, Chile, the United Kingdom, and the United
States of America.


{\it Facilities:} \facility{Spitzer (IRAC)}, \facility{USNO:40in}, \facility{UKIRT}, \facility{NMSU:1m}, \facility{LO:0.8m}, \facility{Hale:5m}, , \facility{Keck:I}, \facility{Keck:II}, \facility{KPNO:Mayall:4m}



\appendix
\section{Additional Objects of Potential Interest}

We have identified two other objects which seem to be eclipsing binaries, but
where the data we collected suggest that the objects are not bona-fide Orion members.
We describe these two objects below.

\subsection{ISOY J0536-0500 }

ISOY J0536-0500 has not been previously 
labeled as an ONC member. It is fainter than all previously known EBs and, because 
it is located further to the NE of the Trapezium, it has not been included in 
most of the studies of the region. Its brightness and colors are very similar 
to those of the known brown-dwarf EB 2M0535-05 as is shown in Figure~\ref{CMD}. 
Its position in the color-magnitude diagram indicated that  ISOY 0536-0500 could 
comprise two substellar objects of even lower mass than those in 2M0535-05. 

ISOY J0536-0500 light curves show several eclipses, both in IRAC [3.6] and [4.5] 
and $J$ bands are clear. This source was also included in the $I_c$ band monitoring 
of \citet{Stassun99} and we have used their $I_c$ band data to obtain a more accurate 
period. The timing analysis gives a period of 3.5705 days and the ephemeris 
we calculate is:

\begin{center}
HJD$_0$=2455128.74 \\
Period=3.570535 $\pm$ 0.00007 days\\
\end{center}

\begin{figure}
\epsscale{.6}
\plotone{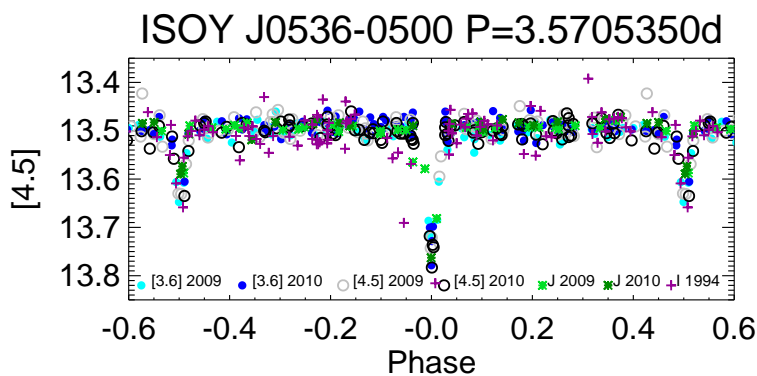}
\caption{Phased light curves for ISOY J0536-0500 folded with a period of 3.5705 days where all the data from 2009 and 2010 are plotted. The symbols are $\bullet$: [3.6] IRAC; $\circ$: [4.5] IRAC;  $*$: $J$ UKIRT, $+$: $I_c$ band from \citet{Stassun99}. [3.6] , $J$, $I_c$ light curves have been shifted in the y-axis to match the mean [4.5] value ([3.6]-[4.5]=0.02, $J$-[4.5]=1.44, I-[4.5]=3.81). \label{43218phased}}
\end{figure}

The phased light curve based on these parameters can be seen in 
Figure~\ref{43218phased}. Again, the $I_c$ band data from 1994 confirms the eclipse features.

The problem with this picture is that the near-IR spectrum which we obtained does
not correspond to that of a late-type dwarf.   Instead, the TSpec spectrum indicates
the star is much hotter, with an inferred spectral type of G2 or earlier (see Figure~\ref{TSpec}).  The
broad-band colors are inconsistent with this spectral type unless the star has
a reddening of order A$_v\sim$ 6.  The location of the star in a CMD (see
Figure~\ref{CMD}) is inconsistent with this reddening unless the object is
not a member of the ONC.  Given the properties of
this star, we suggest that it is a field EB located behind the Orion molecular
cloud and seen through an edge of the cloud.

\subsection{ISOY J0534-0449}

ISOY J0534-0449 was identified by \citet{Stassun99} as the fastest rotator in a study of the period distribution in the ONC \citep[P=0.27 days, Star 1161 of ][]{Stassun99}. \citet{Rebull01} photometrically monitored the outer ONC finding the same period (Star 1450 in their study), and \citet{Carpenter01} finds our EB candidate periodic as well but with a period of 3.14 days (their cadence did not allow them to find shorter periods). 

\begin{figure}
\epsscale{.6}
\plotone{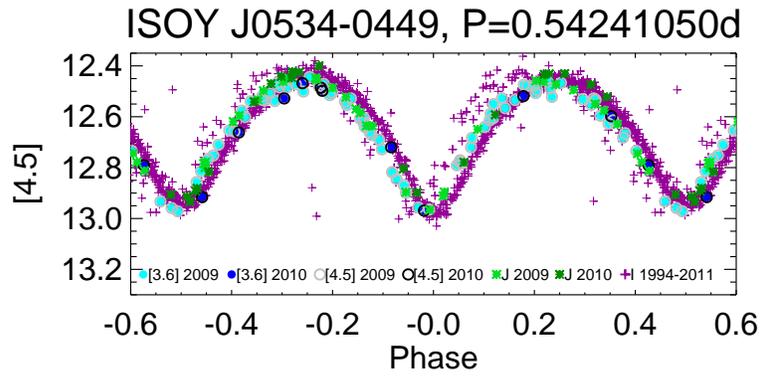}
\caption{Phased light curves for ISOY J0534-0449 folded with a period of 0.542413 days where all our data from 2009 and 2010 are plotted together with 1994-1997 $I_c$ band data from \citet{Stassun99} and \citet{Rebull01}. The symbols are $\bullet$: [3.6] IRAC; $\circ$: [4.5] IRAC; $*$: $J$ UKIRT; and $+$: $I_c$ USNO+SMMV+R01. [3.6] , $J$, and $I_c$ band  light curves have been shifted in the y-axis to match the mean [4.5] value ([3.6]-[4.5]=0.02, $J$-[4.5]=1.41, $I_c$-[4.5]=2.81). \label{2592phased}}
\end{figure}

Figure~\ref{2592phased} includes all the YSOVAR data from 2009 to 2011 plus the $I_c$-band monitoring data from \citet{Stassun99} (obtained on December 1994) and \citet{Rebull01} 
(obtained between December 1995 and February 1997) folded with a period of 0.5424. The light curve shape
appears remarkably similar in all these datasets, spanning 17 years at multiple
epochs.

\begin{figure}
\epsscale{1.}
\plotone{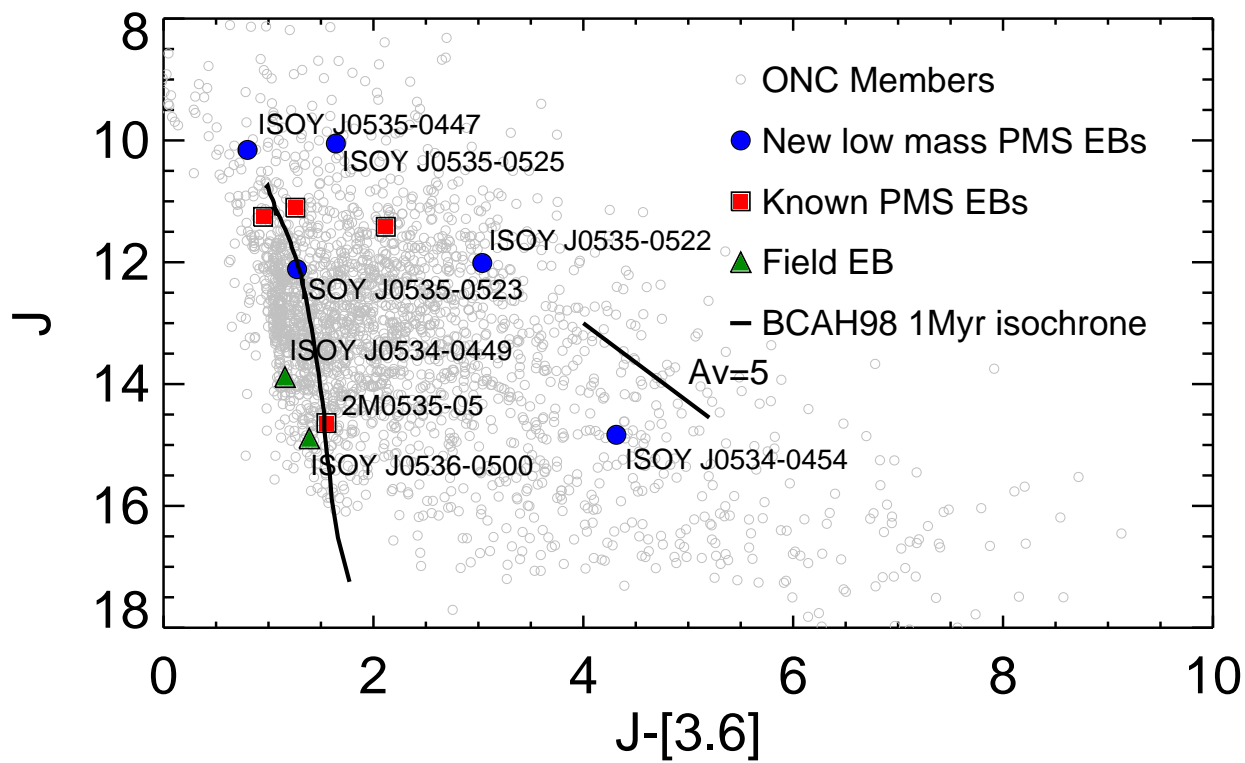}
\caption{$J$ vs. $J$-[3.6]  color-magnitude diagram showing the location of our 6 PMS EBs (blue circles) and the previously known ones (red squares) together with known Orion members (grey circles). The position of the two suspected field EBs is also shown with green triangles. A 1 Myr isochrone is plotted as a solid line. $\theta^1$ Ori E is not shown due to the lack of  $J$-band data.\label{CMD}}
\end{figure}

Its published spectral type is K5 \citep{Rebull01}); however, its location in a $J$ vs. $J-$[3.6] diagram shows it to be much fainter than the other Orion members of similar spectral type - but not significantly reddened (see Figure~\ref{CMD}). In addition, our single epoch HIRES spectrum shows that about the only feature visibile is a fairly strong H$_\alpha$ absorption line instead of in emission as it would be expected for a young mid-K star.
If it were a single, field rapid rotator, presumably
the light variations would be due to large, non-axisymmetrically distributed spots.
However, it seems very unlikely to us that such spots could yield such a stable
light curve shape over such a long period.  Instead, 
we believe the object is more likely to be a field contact binary composed of two $\sim$G dwarfs.





\begin{thebibliography}{73}
\expandafter\ifx\csname natexlab\endcsname\relax\def\natexlab#1{#1}\fi

\bibitem[{{Alecian} {et~al.}(2005){Alecian}, {Catala}, {van't Veer-Menneret},
  {Goupil}, \& {Balona}}]{Alecian05}
{Alecian}, E., {Catala}, C., {van't Veer-Menneret}, C., {Goupil}, M.-J., \&
  {Balona}, L. 2005, \aap, 442, 993

\bibitem[{{Alecian} {et~al.}(2007){Alecian}, {Goupil}, {Lebreton}, {Dupret}, \&
  {Catala}}]{Alecian07}
{Alecian}, E., {Goupil}, M.-J., {Lebreton}, Y., {Dupret}, M.-A., \& {Catala},
  C. 2007, \aap, 465, 241

\bibitem[{{Boden} {et~al.}(2005){Boden}, {Sargent}, {Akeson}, {Carpenter},
  {Torres}, {Latham}, {Soderblom}, {Nelan}, {Franz}, \& {Wasserman}}]{Boden05}
{Boden}, A.~F., {Sargent}, A.~I., {Akeson}, R.~L., {et~al.} 2005, \apj, 635,
  442

\bibitem[{{Cargile} {et~al.}(2008){Cargile}, {Stassun}, \&
  {Mathieu}}]{Cargile08}
{Cargile}, P.~A., {Stassun}, K.~G., \& {Mathieu}, R.~D. 2008, \apj, 674, 329

\bibitem[{{Carpenter} {et~al.}(2001){Carpenter}, {Hillenbrand}, \&
  {Skrutskie}}]{Carpenter01}
{Carpenter}, J.~M., {Hillenbrand}, L.~A., \& {Skrutskie}, M.~F. 2001, \aj, 121,
  3160

\bibitem[{{Casey} {et~al.}(1998){Casey}, {Mathieu}, {Vaz}, {Andersen}, \&
  {Suntzeff}}]{Casey98}
{Casey}, B.~W., {Mathieu}, R.~D., {Vaz}, L.~P.~R., {Andersen}, J., \&
  {Suntzeff}, N.~B. 1998, \aj, 115, 1617

\bibitem[{{Chabrier} {et~al.}(2007){Chabrier}, {Gallardo}, \&
  {Baraffe}}]{Chabrier07}
{Chabrier}, G., {Gallardo}, J., \& {Baraffe}, I. 2007, \aap, 472, L17

\bibitem[{{Cody} \& {Hillenbrand}(2011)}]{Cody11}
{Cody}, A.~M. \& {Hillenbrand}, L.~A. 2011, \apj, 741, 9

\bibitem[{{Costero} {et~al.}(2008){Costero}, {Allen}, {Echevarr{\'{\i}}a},
  {Georgiev}, {Poveda}, \& {Richer}}]{Costero08}
{Costero}, R., {Allen}, C., {Echevarr{\'{\i}}a}, J., {et~al.} 2008, in Revista
  Mexicana de Astronomia y Astrofisica Conference Series, Vol.~34, Revista
  Mexicana de Astronomia y Astrofisica Conference Series, 102--105

\bibitem[{{Costero} {et~al.}(2006){Costero}, {Echevarria}, {Richer}, {Poveda},
  \& {Li}}]{Costero06}
{Costero}, R., {Echevarria}, J., {Richer}, M.~G., {Poveda}, A., \& {Li}, W.
  2006, \iaucirc, 8669, 2

\bibitem[{{Covino} {et~al.}(2000){Covino}, {Catalano}, {Frasca}, {Marilli},
  {Fern{\'a}ndez}, {Alcal{\'a}}, {Melo}, {Paladino}, {Sterzik}, \&
  {Stelzer}}]{Covino00}
{Covino}, E., {Catalano}, S., {Frasca}, A., {et~al.} 2000, \aap, 361, L49

\bibitem[{{Cushing} {et~al.}(2004){Cushing}, {Vacca}, \& {Rayner}}]{Cushing04}
{Cushing}, M.~C., {Vacca}, W.~D., \& {Rayner}, J.~T. 2004, \pasp, 116, 362

\bibitem[{{Cutri} {et~al.}(2003){Cutri}, {Skrutskie}, {van Dyk}, {Beichman},
  {Carpenter}, {Chester}, {Cambresy}, {Evans}, {Fowler}, {Gizis}, {Howard},
  {Huchra}, {Jarrett}, {Kopan}, {Kirkpatrick}, {Light}, {Marsh}, {McCallon},
  {Schneider}, {Stiening}, {Sykes}, {Weinberg}, {Wheaton}, {Wheelock}, \&
  {Zacarias}}]{Cutri03}
{Cutri}, R.~M., {Skrutskie}, M.~F., {van Dyk}, S., {et~al.} 2003, {2MASS All
  Sky Catalog of point sources.} (The IRSA 2MASS All-Sky Point Source Catalog,
  NASA/IPAC Infrared Science
  Archive.~http://irsa.ipac.caltech.edu/applications/Gator/)

\bibitem[{{Da Rio} {et~al.}(2010){Da Rio}, {Robberto}, {Soderblom}, {Panagia},
  {Hillenbrand}, {Palla}, \& {Stassun}}]{DaRio10}
{Da Rio}, N., {Robberto}, M., {Soderblom}, D.~R., {et~al.} 2010, \apj, 722,
  1092

\bibitem[{{Fazio} {et~al.}(2004){Fazio}, {Hora}, {Allen}, {Ashby}, {Barmby},
  {Deutsch}, {Huang}, {Kleiner}, {Marengo}, {Megeath}, {Melnick}, {Pahre},
  {Patten}, {Polizotti}, {Smith}, {Taylor}, {Wang}, {Willner}, {Hoffmann},
  {Pipher}, {Forrest}, {McMurty}, {McCreight}, {McKelvey}, {McMurray}, {Koch},
  {Moseley}, {Arendt}, {Mentzell}, {Marx}, {Losch}, {Mayman}, {Eichhorn},
  {Krebs}, {Jhabvala}, {Gezari}, {Fixsen}, {Flores}, {Shakoorzadeh}, {Jungo},
  {Hakun}, {Workman}, {Karpati}, {Kichak}, {Whitley}, {Mann}, {Tollestrup},
  {Eisenhardt}, {Stern}, {Gorjian}, {Bhattacharya}, {Carey}, {Nelson},
  {Glaccum}, {Lacy}, {Lowrance}, {Laine}, {Reach}, {Stauffer}, {Surace},
  {Wilson}, {Wright}, {Hoffman}, {Domingo}, \& {Cohen}}]{Fazio04}
{Fazio}, G.~G., {Hora}, J.~L., {Allen}, L.~E., {et~al.} 2004, \apjs, 154, 10

\bibitem[{{Feigelson} {et~al.}(2002){Feigelson}, {Broos}, {Gaffney}, {Garmire},
  {Hillenbrand}, {Pravdo}, {Townsley}, \& {Tsuboi}}]{Feigelson02}
{Feigelson}, E.~D., {Broos}, P., {Gaffney}, III, J.~A., {et~al.} 2002, \apj,
  574, 258

\bibitem[{{Felli} {et~al.}(1993{\natexlab{a}}){Felli}, {Churchwell}, {Wilson},
  \& {Taylor}}]{Felli93a}
{Felli}, M., {Churchwell}, E., {Wilson}, T.~L., \& {Taylor}, G.~B.
  1993{\natexlab{a}}, \aaps, 98, 137

\bibitem[{{Felli} {et~al.}(1993{\natexlab{b}}){Felli}, {Taylor}, {Catarzi},
  {Churchwell}, \& {Kurtz}}]{Felli93b}
{Felli}, M., {Taylor}, G.~B., {Catarzi}, M., {Churchwell}, E., \& {Kurtz}, S.
  1993{\natexlab{b}}, \aaps, 101, 127

\bibitem[{{Garcia} {et~al.}(2011){Garcia}, {Stassun}, {Hebb}, {G{\'o}mez Maqueo
  Chew}, \& {Heiser}}]{Garcia11}
{Garcia}, E.~V., {Stassun}, K.~G., {Hebb}, L., {G{\'o}mez Maqueo Chew}, Y., \&
  {Heiser}, A. 2011, \aj, 142, 27

\bibitem[{{Getman} {et~al.}(2005){Getman}, {Feigelson}, {Grosso},
  {McCaughrean}, {Micela}, {Broos}, {Garmire}, \& {Townsley}}]{Getman05}
{Getman}, K.~V., {Feigelson}, E.~D., {Grosso}, N., {et~al.} 2005, \apjs, 160,
  353

\bibitem[{{Gutermuth} {et~al.}(2009){Gutermuth}, {Megeath}, {Myers}, {Allen},
  {Pipher}, \& {Fazio}}]{Gutermuth09}
{Gutermuth}, R.~A., {Megeath}, S.~T., {Myers}, P.~C., {et~al.} 2009, \apjs,
  184, 18

\bibitem[{{Hebb} {et~al.}(2011){Hebb}, {Cegla}, {Stassun}, {Stempels},
  {Cargile}, \& {Palladino}}]{Hebb11}
{Hebb}, L., {Cegla}, H.~M., {Stassun}, K.~G., {et~al.} 2011, \aap, 531, A61+

\bibitem[{{Hebb} {et~al.}(2010){Hebb}, {Stempels}, {Aigrain},
  {Collier-Cameron}, {Hodgkin}, {Irwin}, {Maxted}, {Pollacco}, {Street},
  {Wilson}, \& {Stassun}}]{Hebb10}
{Hebb}, L., {Stempels}, H.~C., {Aigrain}, S., {et~al.} 2010, \aap, 522, A37+

\bibitem[{{Herbig} \& {Griffin}(2006)}]{Herbig06}
{Herbig}, G.~H. \& {Griffin}, R.~F. 2006, \aj, 132, 1763

\bibitem[{{Herbst} {et~al.}(2000){Herbst}, {Rhode}, {Hillenbrand}, \&
  {Curran}}]{Herbst00}
{Herbst}, W., {Rhode}, K.~L., {Hillenbrand}, L.~A., \& {Curran}, G. 2000, \aj,
  119, 261

\bibitem[{{Herter} {et~al.}(2008){Herter}, {Henderson}, {Wilson}, {Matthews},
  {Rahmer}, {Bonati}, {Muirhead}, {Adams}, {Lloyd}, {Skrutskie}, {Moon},
  {Parshley}, {Nelson}, {Martinache}, \& {Gull}}]{Herter08}
{Herter}, T.~L., {Henderson}, C.~P., {Wilson}, J.~C., {et~al.} 2008, in Society
  of Photo-Optical Instrumentation Engineers (SPIE) Conference Series, Vol.
  7014, Society of Photo-Optical Instrumentation Engineers (SPIE) Conference
  Series

\bibitem[{{Hillenbrand}(1997)}]{H97}
{Hillenbrand}, L.~A. 1997, \aj, 113, 1733

\bibitem[{{Hillenbrand} \& {White}(2004)}]{H04}
{Hillenbrand}, L.~A. \& {White}, R.~J. 2004, \apj, 604, 741

\bibitem[{{Irwin} {et~al.}(2007){Irwin}, {Aigrain}, {Hodgkin}, {Stassun},
  {Hebb}, {Irwin}, {Moraux}, {Bouvier}, {Alapini}, {Alexander}, {Bramich},
  {Holtzman}, {Mart{\'{\i}}n}, {McCaughrean}, {Pont}, {Verrier}, \& {Zapatero
  Osorio}}]{Irwin07}
{Irwin}, J., {Aigrain}, S., {Hodgkin}, S., {et~al.} 2007, \mnras, 380, 541

\bibitem[{{Jones} \& {Walker}(1988)}]{Jones88}
{Jones}, B.~F. \& {Walker}, M.~F. 1988, \aj, 95, 1755

\bibitem[{{Kenyon} \& {Hartmann}(1995)}]{Kenyon95}
{Kenyon}, S.~J. \& {Hartmann}, L. 1995, \apjs, 101, 117

\bibitem[{{Kov{\'a}cs} {et~al.}(2002){Kov{\'a}cs}, {Zucker}, \&
  {Mazeh}}]{Kovacs02}
{Kov{\'a}cs}, G., {Zucker}, S., \& {Mazeh}, T. 2002, \aap, 391, 369

\bibitem[{{Kraus} {et~al.}(2011){Kraus}, {Tucker}, {Thompson}, {Craine}, \&
  {Hillenbrand}}]{Kraus11}
{Kraus}, A.~L., {Tucker}, R.~A., {Thompson}, M.~I., {Craine}, E.~R., \&
  {Hillenbrand}, L.~A. 2011, \apj, 728, 48

\bibitem[{{Ku} {et~al.}(1982){Ku}, {Righini-Cohen}, \& {Simon}}]{Ku82}
{Ku}, W.~H.-M., {Righini-Cohen}, G., \& {Simon}, M. 1982, Science, 215, 61

\bibitem[{{Luhman}(1999)}]{Luhman99}
{Luhman}, K.~L. 1999, \apj, 525, 466

\bibitem[{{Luhman} {et~al.}(2000){Luhman}, {Rieke}, {Young}, {Cotera}, {Chen},
  {Rieke}, {Schneider}, \& {Thompson}}]{Luhman00}
{Luhman}, K.~L., {Rieke}, G.~H., {Young}, E.~T., {et~al.} 2000, \apj, 540, 1016

\bibitem[{{Luhman} {et~al.}(2006){Luhman}, {Whitney}, {Meade}, {Babler},
  {Indebetouw}, {Bracker}, \& {Churchwell}}]{Luhman06b}
{Luhman}, K.~L., {Whitney}, B.~A., {Meade}, M.~R., {et~al.} 2006, \apj, 647,
  1180

\bibitem[{{Macdonald} \& {Mullan}(2010)}]{Macdonald10}
{Macdonald}, J. \& {Mullan}, D.~J. 2010, \apj, 723, 1599

\bibitem[{{McLean} {et~al.}(1998){McLean}, {Becklin}, {Bendiksen}, {Brims},
  {Canfield}, {Figer}, {Graham}, {Hare}, {Lacayanga}, {Larkin}, {Larson},
  {Levenson}, {Magnone}, {Teplitz}, \& {Wong}}]{McLean98}
{McLean}, I.~S., {Becklin}, E.~E., {Bendiksen}, O., {et~al.} 1998, in Society
  of Photo-Optical Instrumentation Engineers (SPIE) Conference Series, Vol.
  3354, Society of Photo-Optical Instrumentation Engineers (SPIE) Conference
  Series, ed. {A.~M.~Fowler}, 566--578

\bibitem[{{McLean} {et~al.}(2000){McLean}, {Graham}, {Becklin}, {Figer},
  {Larkin}, {Levenson}, \& {Teplitz}}]{McLean00}
{McLean}, I.~S., {Graham}, J.~R., {Becklin}, E.~E., {et~al.} 2000, in Society
  of Photo-Optical Instrumentation Engineers (SPIE) Conference Series, Vol.
  4008, Society of Photo-Optical Instrumentation Engineers (SPIE) Conference
  Series, ed. {M.~Iye \& A.~F.~Moorwood}, 1048--1055

\bibitem[{{McNamara}(1976)}]{McNamara76}
{McNamara}, B.~J. 1976, \aj, 81, 845

\bibitem[{{Mohanty} {et~al.}(2010){Mohanty}, {Stassun}, \&
  {Doppmann}}]{Mohanty10}
{Mohanty}, S., {Stassun}, K.~G., \& {Doppmann}, G.~W. 2010, \apj, 722, 1138

\bibitem[{{Morales} {et~al.}(2010){Morales}, {Gallardo}, {Ribas}, {Jordi},
  {Baraffe}, \& {Chabrier}}]{MoralesJC10}
{Morales}, J.~C., {Gallardo}, J., {Ribas}, I., {et~al.} 2010, \apj, 718, 502

\bibitem[{{Morales-Calder{\'o}n} {et~al.}(2011){Morales-Calder{\'o}n},
  {Stauffer}, {Hillenbrand}, {Gutermuth}, {Song}, {Rebull}, {Plavchan},
  {Carpenter}, {Whitney}, {Covey}, {Alves de Oliveira}, {Winston},
  {McCaughrean}, {Bouvier}, {Guieu}, {Vrba}, {Holtzman}, {Marchis}, {Hora},
  {Wasserman}, {Terebey}, {Megeath}, {Guinan}, {Forbrich}, {Hu{\'e}lamo},
  {Riviere-Marichalar}, {Barrado}, {Stapelfeldt}, {Hern{\'a}ndez}, {Allen},
  {Ardila}, {Bayo}, {Favata}, {James}, {Werner}, \& {Wood}}]{Morales11}
{Morales-Calder{\'o}n}, M., {Stauffer}, J.~R., {Hillenbrand}, L.~A., {et~al.}
  2011, ArXiv e-prints

\bibitem[{{Morales-Calder{\'o}n} {et~al.}(2006){Morales-Calder{\'o}n},
  {Stauffer}, {Kirkpatrick}, {Carey}, {Gelino}, {Barrado y Navascu{\'e}s},
  {Rebull}, {Lowrance}, {Marley}, {Charbonneau}, {Patten}, {Megeath}, \&
  {Buzasi}}]{Morales06}
{Morales-Calder{\'o}n}, M., {Stauffer}, J.~R., {Kirkpatrick}, J.~D., {et~al.}
  2006, \apj, 653, 1454

\bibitem[{{Parenago}(1954)}]{Parenago54}
{Parenago}, P.~P. 1954, Trudy Gosudarstvennogo Astronomicheskogo Instituta, 25,
  1

\bibitem[{{Popper}(1987)}]{Popper87}
{Popper}, D.~M. 1987, \apjl, 313, L81

\bibitem[{{Prato}(2007)}]{Prato07}
{Prato}, L. 2007, \apj, 657, 338

\bibitem[{{Prsa} \& {Zwitter}(2005)}]{Phoebe05}
{Prsa}, A. \& {Zwitter}, T. 2005, in ESA Special Publication, Vol. 576, The
  Three-Dimensional Universe with Gaia, ed. {C.~Turon, K.~S.~O'Flaherty, \&
  M.~A.~C.~Perryman}, 611

\bibitem[{{Ram{\'{\i}}rez} {et~al.}(2004){Ram{\'{\i}}rez}, {Rebull},
  {Stauffer}, {Strom}, {Hillenbrand}, {Hearty}, {Kopan}, {Pravdo}, {Makidon},
  \& {Jones}}]{Ramirez04}
{Ram{\'{\i}}rez}, S.~V., {Rebull}, L., {Stauffer}, J., {et~al.} 2004, \aj, 128,
  787

\bibitem[{{Rayner} {et~al.}(1989){Rayner}, {McLean}, {Aspin}, \&
  {McCaughrean}}]{Rayner89}
{Rayner}, J., {McLean}, I., {Aspin}, C., \& {McCaughrean}, M. 1989, \mnras,
  241, 469

\bibitem[{{Rayner} {et~al.}(2009){Rayner}, {Cushing}, \& {Vacca}}]{Rayner09}
{Rayner}, J.~T., {Cushing}, M.~C., \& {Vacca}, W.~D. 2009, \apjs, 185, 289

\bibitem[{{Rebull}(2001)}]{Rebull01}
{Rebull}, L.~M. 2001, \aj, 121, 1676

\bibitem[{{Reiners} {et~al.}(2007){Reiners}, {Seifahrt}, {Stassun}, {Melo}, \&
  {Mathieu}}]{Reiners07}
{Reiners}, A., {Seifahrt}, A., {Stassun}, K.~G., {Melo}, C., \& {Mathieu},
  R.~D. 2007, \apjl, 671, L149

\bibitem[{{Sicilia-Aguilar} {et~al.}(2005){Sicilia-Aguilar}, {Hartmann},
  {Hern{\'a}ndez}, {Brice{\~n}o}, \& {Calvet}}]{Sicilia05}
{Sicilia-Aguilar}, A., {Hartmann}, L.~W., {Hern{\'a}ndez}, J., {Brice{\~n}o},
  C., \& {Calvet}, N. 2005, \aj, 130, 188

\bibitem[{{Siess} {et~al.}(2000){Siess}, {Dufour}, \& {Forestini}}]{Siess00}
{Siess}, L., {Dufour}, E., \& {Forestini}, M. 2000, \aap, 358, 593

\bibitem[{{Simon} {et~al.}(2000){Simon}, {Dutrey}, \& {Guilloteau}}]{Simon00}
{Simon}, M., {Dutrey}, A., \& {Guilloteau}, S. 2000, \apj, 545, 1034

\bibitem[{{Stassun} {et~al.}(2008){Stassun}, {Mathieu}, {Cargile}, {Aarnio},
  {Stempels}, \& {Geller}}]{Stassun08}
{Stassun}, K.~G., {Mathieu}, R.~D., {Cargile}, P.~A., {et~al.} 2008, \nat, 453,
  1079

\bibitem[{{Stassun} {et~al.}(1999){Stassun}, {Mathieu}, {Mazeh}, \&
  {Vrba}}]{Stassun99}
{Stassun}, K.~G., {Mathieu}, R.~D., {Mazeh}, T., \& {Vrba}, F.~J. 1999, \aj,
  117, 2941

\bibitem[{{Stassun} {et~al.}(2006){Stassun}, {Mathieu}, \&
  {Valenti}}]{Stassun06}
{Stassun}, K.~G., {Mathieu}, R.~D., \& {Valenti}, J.~A. 2006, \nat, 440, 311

\bibitem[{{Stassun} {et~al.}(2007){Stassun}, {Mathieu}, \&
  {Valenti}}]{Stassun07}
{Stassun}, K.~G., {Mathieu}, R.~D., \& {Valenti}, J.~A. 2007, \apj, 664, 1154

\bibitem[{{Stassun} {et~al.}(2004){Stassun}, {Mathieu}, {Vaz}, {Stroud}, \&
  {Vrba}}]{Stassun04}
{Stassun}, K.~G., {Mathieu}, R.~D., {Vaz}, L.~P.~R., {Stroud}, N., \& {Vrba},
  F.~J. 2004, \apjs, 151, 357

\bibitem[{{Stelzer} {et~al.}(2005){Stelzer}, {Flaccomio}, {Montmerle},
  {Micela}, {Sciortino}, {Favata}, {Preibisch}, \& {Feigelson}}]{Stelzer05}
{Stelzer}, B., {Flaccomio}, E., {Montmerle}, T., {et~al.} 2005, \apjs, 160, 557

\bibitem[{{Stempels} {et~al.}(2008){Stempels}, {Hebb}, {Stassun}, {Holtzman},
  {Dunstone}, {Glowienka}, \& {Frandsen}}]{Stempels08}
{Stempels}, H.~C., {Hebb}, L., {Stassun}, K.~G., {et~al.} 2008, \aap, 481, 747

\bibitem[{{Tian} {et~al.}(1996){Tian}, {van Leeuwen}, {Zhao}, \& {Su}}]{Tian96}
{Tian}, K.~P., {van Leeuwen}, F., {Zhao}, J.~L., \& {Su}, C.~G. 1996, \aaps,
  118, 503

\bibitem[{{Tobin} {et~al.}(2009){Tobin}, {Hartmann}, {Furesz}, {Mateo}, \&
  {Megeath}}]{Tobin09}
{Tobin}, J.~J., {Hartmann}, L., {Furesz}, G., {Mateo}, M., \& {Megeath}, S.~T.
  2009, \apj, 697, 1103

\bibitem[{{Tognelli} {et~al.}(2011){Tognelli}, {Prada Moroni}, \&
  {Degl'Innocenti}}]{Tognelli11}
{Tognelli}, E., {Prada Moroni}, P.~G., \& {Degl'Innocenti}, S. 2011, \aap, 533,
  A109

\bibitem[{{Vacca} {et~al.}(2004){Vacca}, {Cushing}, \& {Rayner}}]{Vacca04}
{Vacca}, W.~D., {Cushing}, M.~C., \& {Rayner}, J.~T. 2004, \pasp, 116, 352

\bibitem[{{van Hamme}(1993)}]{vanHamme93}
{van Hamme}, W. 1993, \aj, 106, 2096

\bibitem[{{Vogt} {et~al.}(1994){Vogt}, {Allen}, {Bigelow}, {Bresee}, {Brown},
  {Cantrall}, {Conrad}, {Couture}, {Delaney}, {Epps}, {Hilyard}, {Hilyard},
  {Horn}, {Jern}, {Kanto}, {Keane}, {Kibrick}, {Lewis}, {Osborne},
  {Pardeilhan}, {Pfister}, {Ricketts}, {Robinson}, {Stover}, {Tucker}, {Ward},
  \& {Wei}}]{Vogt94}
{Vogt}, S.~S., {Allen}, S.~L., {Bigelow}, B.~C., {et~al.} 1994, in Society of
  Photo-Optical Instrumentation Engineers (SPIE) Conference Series, Vol. 2198,
  Society of Photo-Optical Instrumentation Engineers (SPIE) Conference Series,
  ed. {D.~L.~Crawford \& E.~R.~Craine}, 362

\bibitem[{{Werner} {et~al.}(2004){Werner}, {Roellig}, {Low}, {Rieke}, {Rieke},
  {Hoffmann}, {Young}, {Houck}, {Brandl}, {Fazio}, {Hora}, {Gehrz}, {Helou},
  {Soifer}, {Stauffer}, {Keene}, {Eisenhardt}, {Gallagher}, {Gautier}, {Irace},
  {Lawrence}, {Simmons}, {Van Cleve}, {Jura}, {Wright}, \&
  {Cruikshank}}]{Werner04}
{Werner}, M.~W., {Roellig}, T.~L., {Low}, F.~J., {et~al.} 2004, \apjs, 154, 1

\bibitem[{{Wilson} {et~al.}(2004){Wilson}, {Henderson}, {Herter}, {Matthews},
  {Skrutskie}, {Adams}, {Moon}, {Smith}, {Gautier}, {Ressler}, {Soifer}, {Lin},
  {Howard}, {LaMarr}, {Stolberg}, \& {Zink}}]{Wilson04}
{Wilson}, J.~C., {Henderson}, C.~P., {Herter}, T.~L., {et~al.} 2004, in Society
  of Photo-Optical Instrumentation Engineers (SPIE) Conference Series, Vol.
  5492, Society of Photo-Optical Instrumentation Engineers (SPIE) Conference
  Series, ed. {A.~F.~M.~Moorwood \& M.~Iye}, 1295--1305

\bibitem[{{Wilson} \& {Devinney}(1971)}]{WD71}
{Wilson}, R.~E. \& {Devinney}, E.~J. 1971, \apj, 166, 605

\end{thebibliography}

\clearpage



\clearpage

\begin{deluxetable}{rcccccccccccc}
\rotate
\tabletypesize{\tiny}
\tablecaption{New eclipsing binary candidates}
\tablewidth{0pt}
\tablehead{
\colhead{Source\tablenotemark{a}} & \colhead{Most Common Name\tablenotemark{b}} &  \colhead{V (mag)}&  \colhead{I$_c$ (mag)}&  \colhead{J (mag)\tablenotemark{e}}&  \colhead{H (mag)\tablenotemark{e}}&  \colhead{K$_s$ (mag)\tablenotemark{e}}&  \colhead{[3.6] (mag)\tablenotemark{f}}&  \colhead{[4.5] (mag)\tablenotemark{f}}&  \colhead{[5.8] (mag)\tablenotemark{f}}&  \colhead{[8.0] (mag)\tablenotemark{f}}&  \colhead{SpT\tablenotemark{h}}&\colhead{Period (days)\tablenotemark{h}}\label{EBperiods}}
\startdata
ISOY J053454.31-045413.0   & 	[RRS2004] NOFF W015 &--&-- &14.83 &12.43 & 11.29& 10.52&10.31 &9.94 &9.26 &M4-M5 &5.1993 \\	
ISOY J053505.71-052354.1   & JW 276 &17.00\tablenotemark{d}&13.96\tablenotemark{d} &12.11 &11.46 &11.16 &10.84 & 10.79&-- &-- &M5-M6 &20.485\\ 
ISOY J053515.55-052514.1 &Parenago 1872 &13.83\tablenotemark{d} &11.55\tablenotemark{d} & 10.05 & 9.11 &  8.74 & 8.41 &8.37 & 7.50 & -- & K4-M1\tablenotemark{d} & -- \\
ISOY J053515.76-052309.9   & theta1 Ori E  & --&--&6.64 &6.24 &6.06 & 6.94&6.52 &6.12 &-- & G2 IV\tablenotemark{g}&9.89520\\
ISOY J053518.03-052205.4   & [H97b] 9209  & 19.06\tablenotemark{d}&15.30\tablenotemark{d}& --&10.36 &9.30 &8.98 &8.67 &-- &-- & K0-K3&5.6175\\
ISOY J053526.88-044730.7   & Parenago 2017 &12.17\tablenotemark{c}&11.06\tablenotemark{c} &10.16 & 9.59&9.42 &9.36 & 9.32& 9.30 & 9.26&K0-K2&3.905625\\
\hline
\hline
ISOY J053446.01-044922.1\tablenotemark{i}   & V1448 Ori & 17.43\tablenotemark{c} &15.61\tablenotemark{c}& 13.88& 13.26&12.95& 12.73& 12.66& 12.91& --& K5\tablenotemark{c} &0.5424125\\
ISOY J053605.95-050041.2\tablenotemark{i} & 2MASS J05360595-0500413& --&--&14.89 &14.09 &13.73 &13.51 &13.49 &13.20 & --& $<$G& 3.570535\\
\hline
\enddata
\label{binaries}
\tablenotetext{a}{Names are composed of an acronym, ISOY (Initial Spitzer Orion YSO), followed by the coordinates of the source.}
\tablenotetext{b}{Most common names from the literature searchable in SIMBAD database: [H97b]: \citet{H97}, JW: \citet{Jones88}, [RRS2004] : \citet{Ramirez04}, Parenago: \citet{Parenago54}}
\tablenotetext{c}{from \citet{Rebull01}}
\tablenotetext{d}{from \citet{H97}}
\tablenotetext{e}{from \citet{Carpenter01}}
\tablenotetext{f}{from Megeath et al. 2012 in prep.}
\tablenotetext{g}{from \citet{Costero06}}
\tablenotetext{h}{from this work}
\tablenotetext{i}{Probable field EBs}
\end{deluxetable}

\begin{deluxetable}{lcccc}
\tabletypesize{\scriptsize}
\tablecaption{Summary of time-series photometric observations}
\tablewidth{0pt}
\tablehead{
\colhead{Source} & \colhead{Data Source\tablenotemark{a}} & \colhead{Band}& \colhead{Date range}& \colhead{\# epochs}}
\startdata
ISOY J053454.31-045413.0   &  Spitzer/IRAC   & [3.6], [4.5] & 23 Oct 2009 - 30 Nov 2011    & 162, 162     \\
				       & Ukirt/WFCAM   & $J$                 & 20 Oct 2009 - 15 Dec 2010 & 54    \\

\hline
ISOY J053505.71-052354.1   &  Spitzer/IRAC   & [3.6], [4.5] & 23 Oct 2009 - 9 Dec 2010    & 166, 166     \\
			       &  APO/31$\arcmin$           & $I_c$                 & 24 Oct 2009 - 30 Nov 2010  &  22 \\
					       & WFI/	 ESO2.1m & $I_c$                 &    21 Nov 2010 - 29 Nov 2010              & 9      \\
					       & LOWELL           & $I_c$                  & 24 Oct 2009 - 1 Dec 2009  &  9 \\
					       & USNO                & $I_c$                 &30 Oct 2010 - 28 Dec 2010   & 2387 \\
					       & SMMV		& $I_c$                 & 13 Dec 1994 - 26 Nov 2002 & 298 \\                                                  					       	       &MP		& $I_c$                 &  17 Nov 2004 - 8 Dec 2006 & 1418 \\                                                         
\hline
ISOY J053515.55-052514.1 & Spitzer/IRAC   & [3.6], [4.5] & 23 Oct 2009 - 9 Dec 2010    & 166, 166     \\
				       & Ukirt/WFCAM   & $J$                 & 20 Oct 2009 - 15 Dec 2010 & 44    \\
			       &  APO/31"           & $I_c$                 & 24 Oct 2009 - 14 Dec 2010  &  69 \\
					       & LOWELL           & $I_c$                  & 24 Oct 2009 - 3 Dec 2009  &  21 \\
					       & USNO                & $I_c$                 &30 Oct 2010 - 28 Dec 2010   & 1933 \\
\hline
$\theta$1 Ori E   		      &  Spitzer/IRAC   & [3.6], [4.5] & 23 Oct 2009 - 9 Dec 2010    & 66, 166     \\
\hline
ISOY J053518.03-052205.4   & Spitzer/IRAC   &  [3.6], [4.5] & 23 Oct 2009 - 9 Dec 2010    & 166, 137     \\
			       & WFI			& $I_c$                 & 21 Nov 2010 - 28 Nov 2010 &   7    \\
\hline
ISOY J053526.88-044730.7   & Spitzer/IRAC   &  [3.6], [4.5] & 23 Oct 2009 - 27 Apr 2011    & 108, 2821     \\
          				       & Ukirt/WFCAM   & $J$                 & 20 Oct 2009 - 15 Dec 2010 & 54    \\
					       & USNO		&  $I_c$                  & 29 Oct 2010 - 25 Feb 2011 & 14927\tablenotemark{c}\\
					       &R01			&  $I_c$		     &10 Dec 1995 - 3 Feb 1997   & 79 \\	
\hline
\hline
ISOY J053446.01-044922.1\tablenotemark{b}   &  Spitzer/IRAC   & [3.6], [4.5]& 23 Oct 2009 - 9 Dec 2010    & 92, 92     \\
				       & Ukirt/WFCAM   & $J$                 & 20 Oct 2009 - 15 Dec 2010 & 54    \\
					       &USNO		&  $I_c$                  & 29 Oct 2010 - 25 Feb 2011 & 1357\\
					       &R01			&  $I_c$		     &10 Dec 1995 - 3 Feb 1997   & 79 \\
					       &SMMV		&  $I_c$	     &13 Dec 1994 - 22 Dec 1994 & 126 \\
\hline
ISOY\_J053605.95-050041.2\tablenotemark{b} &  Spitzer/IRAC   & [3.6], [4.5] & 23 Oct 2009 - 30 Nov 2011    &  178, 178  \\
					       & Ukirt/WFCAM   & $J$                 & 20 Oct 2009 - 15 Dec 2010 & 54 \\   					       &SMMV		&  $I_c$	     &13 Dec 1994 - 22 Dec 1994 & 89 \\
\\
\hline
\enddata
\label{ObsSummary}
\tablenotetext{a}{R01: \citet{Rebull01}; SMMV: \citet{Stassun99}; MP: Monitor project, private communication.}
\tablenotetext{b}{Probable field EBs}
\tablenotetext{c}{The USNO data for ISOY J0535-0447 was binned every 9 datapoints yielding a total of 1624 datapoints}
\end{deluxetable}

\begin{deluxetable}{rccc}
\tabletypesize{\small}
\tablecaption{Radial velocity for ISOY J0535-0447}
\tablewidth{0pt}
\tablehead{
\colhead{HJD} & \colhead{RV (km/s)}& \colhead{eRV (km/s)} & \colhead{Instrument/Telescope}
}
\startdata
2455463.989        &   30. &        2.  & ECHL/KPNO4m\\	
2455522.83089 \tablenotemark{a}  &  23.5 &       2.& NIRSPEC/Keck II\\
2455523.06103 \tablenotemark{a}   & 23.5    &    2. & NIRSPEC/Keck II\\
2455527.05545 &    35.3    &    2.  & NIRSPEC/Keck II \\
2455543.903      &     30.6    &    2. & HIRES/Keck I		\\
2455584.93613  &   27.3   &     2. & NIRSPEC/Keck II		\\
2455597.9375      &    38.   &      2. & NIRSPEC/Keck II		\\
2455635.726    &       28.0   &     2. & HIRES/Keck I		\\
\hline
\enddata
\label{spectable}
\tablenotetext{a}{this epoch was not used for the fit}
\end{deluxetable}

\begin{deluxetable}{rc}
\tabletypesize{\small}
\tablecaption{Orbital parameters of $\theta^1$ Ori E}
\tablewidth{0pt}
\tablehead{
\colhead{Parameter} & \colhead{Value}
}
\startdata
Period & 9.89520 $\pm$ 0.0007 \tablenotemark{a}\\
HJD$_0$ &2453285.98828\tablenotemark{a}\\
a$sin$i & 33.046 $\pm$ 0.106 R$_\odot$\\
i & 73.7 $\pm$ 0.9 deg\\
a & 34.430 $\pm$ 0.193 R$_\odot$\\
q = M$_2$/M$_1$ & 0.9965 $\pm$ 0.0065\\
M1 & 2.807 $\pm$ 0.048 M$_\odot$\\
M2 & 2.797 $\pm$ 0.048 M$_\odot$\\
T$_2$/T$_1$ & 1.12 $\pm$ 0.08 \tablenotemark{b}\\
R$_1$+R$_2$ & 12.5 $\pm$ 0.6 R$_\odot$ \\
\hline
\enddata
\label{ThetaOri_fit}
\tablenotetext{a}{\citet{Costero08}}
\tablenotetext{b}{assumed T$_1$ = 6000 K based on spectral type}
\end{deluxetable}

\begin{deluxetable}{rc}
\tabletypesize{\small}
\tablecaption{Orbital parameters of ISOY J0535-0447 }
\tablewidth{0pt}
\tablehead{
\colhead{Parameter} & \colhead{Value}
}
\startdata
Period & 3.905625 $\pm$0.000030\\
HJD$_0$ &2455126.26\\
i  &88.8 $\pm$ 0.9 deg\\
a &10.0 R$_\odot$\tablenotemark{a}\\
V$_\gamma$ &30.4 km/s\\
q = M$_2$/M$_1$ &0.06\\
M$_1$&  0.83 M$_\odot$\\
M$_2$ &0.05 M$_\odot$\\
T$_2$/T$_1$ &0.55 $\pm$ 0.03  \tablenotemark{b}\\
R$_1$ + R$_2$ & 2.87 $\pm$ 0.03 R$_\odot$\\
\hline
\enddata
\label{EB3227_fit}
\tablenotetext{a}{The semi-major axis was fixed, and therefore the mass ratio and the individual masses are estimated}
\tablenotetext{b}{T$_1$ = 5150 was adopted from K0 spectral type}
\end{deluxetable}

\begin{deluxetable}{rccc}
\tabletypesize{\small}
\tablecaption{$\theta^1$ Ori E time series at [3.6] and [4.5]. (Full table
available in the electronic version of this article.)}
\tablewidth{0pt}
\tablehead{
\colhead{HJD}&\colhead{Filter} &\colhead{Mag} & \colhead{Error}
}
\startdata
  2455128.92141       &                         IRAC1&     6.793&     0.027\\
  2455129.30254          &                      IRAC1   &  6.802   &  0.026\\
  2455130.43992             &                   IRAC1    & 6.807    & 0.027\\
  2455131.24015                &                IRAC1     &6.828     &0.030 \\
  2455132.78887                   &             IRAC1     &6.803     &0.028\\
  2455133.3197                        &         IRAC1     &6.806     &0.027\\
  2455135.00469                        &        IRAC1     &6.769     &0.030 \\
  2455135.67994                           &     IRAC1     &6.779     &0.030 \\
  2455137.19755                              &  IRAC1     &6.776     &0.027\\
  2455140.00532                                &IRAC1     &6.785     &0.028\\
\hline
\enddata
\label{EB45099_timeseries}
\end{deluxetable}

\begin{deluxetable}{rccc}
\tabletypesize{\small}
\tablecaption{ ISOY J0535-0447 time series at [3.6], [4.5], J, and $Ic$ bands. (Full table
available in the electronic version of this article.)}
\tablewidth{0pt}
\tablehead{
\colhead{HJD}&\colhead{Filter} &\colhead{Mag} & \colhead{Error}
}
\startdata
  2455128.00636     & IRAC1 &     9.273 &0.003 \\
  2455128.26383     & IRAC1 &     9.309 &0.003 \\
  2455128.50885     & IRAC1 &     9.261 &0.003 \\
  2455128.87874     & IRAC1 &     9.267 &0.003 \\
  2455129.25986     & IRAC1 &     9.288 &0.003 \\
  2455129.75754     & IRAC1 &     9.204 &0.003 \\
  2455130.39723     & IRAC1 &     9.245 &0.003 \\
  2455131.19746     & IRAC1 &     9.258 &0.003 \\
  2455131.85941     & IRAC1 &     9.258 &0.003 \\
  2455132.08532     & IRAC1 &     9.300 &0.003 \\
\hline
\enddata
\label{EB3227_timeseriesI}
\end{deluxetable}

\begin{deluxetable}{rccc}
\tabletypesize{\small}
\tablecaption{ ISOY J0535-0522 time series at [3.6], [4.5], and $Ic$ bands. (Full table
available in the electronic version of this article.)}
\tablewidth{0pt}
\tablehead{
\colhead{HJD}&\colhead{Filter} &\colhead{Mag} & \colhead{Error}
}
\startdata
  2455128.04907& IRAC1 &     8.798  &  0.092\\
  2455128.30655& IRAC1 &     8.827  &  0.093\\
  2455128.55158& IRAC1 &     8.845  &  0.097\\
  2455128.92141& IRAC1 &     8.926  &  0.101\\
  2455129.30254& IRAC1 &     8.807  &  0.091\\
  2455129.80022& IRAC1 &     8.826  &  0.092\\
  2455130.43992& IRAC1 &     8.828  &  0.097\\
  2455131.24015& IRAC1 &     8.83   &  0.097\\
  2455131.90210 & IRAC1&     8.822  &  0.099\\
  2455132.12801& IRAC1 &     8.822  &  0.096\\

\hline
\enddata
\label{EB46222_timeseriesI1}
\end{deluxetable}

\begin{deluxetable}{rccc}
\tabletypesize{\small}
\tablecaption{ ISOY J0535-0523 time series at [3.6], [4.5], and $Ic$ bands. (Full table
available in the electronic version of this article.)}
\tablewidth{0pt}
\tablehead{
\colhead{HJD}&\colhead{Filter} &\colhead{Mag} & \colhead{Error}
}
\startdata
  2455128.04658     & IRAC1 &     10.672& 0.044\\
  2455128.30406     & IRAC1 &     10.671& 0.044\\
  2455128.54909     & IRAC1 &     10.653& 0.042\\
  2455128.91892     & IRAC1 &     10.681& 0.043\\
  2455129.30005     & IRAC1 &     10.669& 0.044\\
  2455129.79774     & IRAC1 &     10.650& 0.045\\
  2455130.43743     & IRAC1 &     10.688& 0.045\\
  2455131.23766     & IRAC1 &     10.662& 0.045\\
  2455131.89961     & IRAC1 &     10.667& 0.044\\
  2455132.12552     & IRAC1 &     10.658& 0.042\\
\hline
\enddata
\label{EB40134_timeseriesI1}
\end{deluxetable}

\begin{deluxetable}{rccc}
\tabletypesize{\small}
\tablecaption{ ISOY J0534-0454 time series at [3.6], [4.5], and $Ic$ bands. (Full table
available in the electronic version of this article.)}
\tablewidth{0pt}
\tablehead{
\colhead{HJD}&\colhead{Filter} &\colhead{Mag} & \colhead{Error}
}
\startdata
  2455128.00073 & IRAC1 & 10.472 & 0.039\\
  2455128.25819 & IRAC1 &10.462  &0.038	\\
  2455128.50321 & IRAC1 &10.471  &0.039	\\
  2455128.87311 & IRAC1 &10.451  &0.037	\\
  2455129.25424 & IRAC1 &10.469  &0.039	\\
  2455129.75191 & IRAC1 &10.452  &0.039	\\
  2455130.39161 & IRAC1 &10.473  &0.036	\\
  2455131.19183 & IRAC1 &10.480  &0.039	\\
  2455131.85379 & IRAC1 &10.466  &0.039	\\
  2455132.07970 & IRAC1 &10.460  &0.037	\\

\hline
\enddata
\label{EB1512_timeseriesI1}
\end{deluxetable}

\begin{deluxetable}{rccc}
\tabletypesize{\small}
\tablecaption{ ISOY J0535-0525 time series at [3.6], [4.5], and $Ic$ bands. (Full table
available in the electronic version of this article.)}
\tablewidth{0pt}
\tablehead{
\colhead{HJD}&\colhead{Filter} &\colhead{Mag} & \colhead{Error}
}
\startdata
  2455128.04668& IRAC1 &     8.337 &    0.033  \\
  2455128.30416& IRAC1 &     8.358 &    0.032  \\
  2455128.54919& IRAC1 &     8.380 &     0.033 \\
  2455128.91902& IRAC1 &     8.432 &    0.031  \\
  2455129.30015& IRAC1 &     8.456 &    0.037  \\
  2455129.79818& IRAC1 &     8.494 &    0.036  \\
  2455130.43769& IRAC1 &     8.498 &    0.036  \\
  2455131.23792& IRAC1 &     8.443 &    0.037  \\
  2455131.89987& IRAC1 &     8.389 &    0.035  \\
  2455132.12578& IRAC1 &     8.365 &    0.033  \\
\hline
\enddata
\label{EB40003_timeseriesI1}
\end{deluxetable}




%



\end{document}